\documentclass[trackchanges, twocolumn]{aastex701}

\usepackage{placeins}
\usepackage{xcolor}
\usepackage{amsmath}
\usepackage{graphicx}
\usepackage{comment}
\usepackage{booktabs}
\usepackage{siunitx}
\usepackage{placeins}
\usepackage{float}

\usepackage{subcaption}  

\newcommand{\isotope}[2][]{$^{#1}$#2}

\begin{document}

\title{Kilonova Signatures of Nuclear Mass Model and Composition Variations}

\author[orcid=0009-0005-8103-5823, gname=Alex, sname=Ross]{Alex Ross}
\altaffiliation{A.R. and A.I. are co-first authors, with name ordering decided by coin flip.}
\affiliation{Department of Physics, University of Washington, Seattle, WA, 98195, USA}
\affiliation{Department of Astronomy, University of Washington, Seattle, WA, 98195, USA}
\affiliation{X Computational Physics Division, Los Alamos National Laboratory, Los Alamos, NM 87545, USA}
\email[show]{adr55@uw.edu}
\affiliation{Center for Theoretical Astrophysics, Los Alamos National Laboratory, Los Alamos, NM, 87545, USA}

\author[orcid=0009-0007-0372-9576,gname=Andrew, sname=Imai]{Andrew Imai}
\altaffiliation{A.R. and A.I. are co-first authors, with name ordering decided by coin flip.}
\affiliation{X Computational Physics Division, Los Alamos National Laboratory, Los Alamos, NM 87545, USA}
\affiliation{Center for Theoretical Astrophysics, Los Alamos National Laboratory, Los Alamos, NM, 87545, USA}
\affiliation{Department of Physics and Astronomy, University of Notre Dame, Notre Dame, IN, 46556, USA}
\email[show]{aimai@nd.edu}

\author[orcid=0000-0001-7042-4472,gname=Marko, sname=Ristić]{Marko Ristić} 
\affiliation{Center for Theoretical Astrophysics, Los Alamos National Laboratory, Los Alamos, NM, 87545, USA}
\affiliation{T-5 Applied Mathematics and Plasma Physics Group, Los Alamos National Laboratory, Los Alamos, NM 87545, USA}
\affiliation{Department of Astronomy and Astrophysics, University of California San Diego, La Jolla, CA 92093, USA}
\email{markoris@ucsd.edu}

\author[orcid=0000-0002-9950-9688, gname=Matthew, sname=Mumpower]{Matthew R. Mumpower}
\affiliation{Center for Theoretical Astrophysics, Los Alamos National Laboratory, Los Alamos, NM, 87545, USA}
\affiliation{Department of Physics and Astronomy, University of Notre Dame, Notre Dame, IN, 46556, USA}
\affiliation{Obsidian Research, Fort Wayne, IN, 46835, USA}
\email{matthew@mumpower.net}
\affiliation{Computational Physics and Methods, Los Alamos National Laboratory, Los Alamos, NM, 87545, USA }

\author[orcid=0000-0003-3265-4079,gname=Ryan, sname=Wollaeger]{Ryan Wollaeger}
\affiliation{Center for Theoretical Astrophysics, Los Alamos National Laboratory, Los Alamos, NM, 87545, USA}
\affiliation{Computational Physics and Methods, Los Alamos National Laboratory, Los Alamos, NM, 87545, USA }
\email{wollaeger@lanl.gov}

\begin{abstract}

Neutron star mergers are expected to be one of the sites of heavier-than-iron element nucleosynthesis in the universe through the rapid neutron capture process ($r$-process) in the ejected material. The elements produced are theorized to depend on the initial composition of the ejecta from such a merger, particularly the electron fraction $Y_e$, and on the nuclear mass model used. 
We study how variations in lanthanide, actinide, and heavier element ($Z\geq100$) creation across electron fraction and nuclear mass model variations propagate into simulated kilonova light curves and spectra. We use an established set of ejecta tracers from GRMHD simulations
as input to a nuclear reaction network to simulate the $r$-process, compute heating rates from the resulting nucleosynthesis, and provide the resulting elemental compositions to a Monte Carlo radiative transfer simulation code and generate light curves and spectra for each of our $Y_e$ and nuclear mass model variations. We show that an excess of elements with $Z\geq100$ in certain mass models drives differences in bolometric light curves and spectra through spontaneous fission heating. The strongest observable differences for $Y_e$ variations appear in the multiband color evolution. For nuclear mass model variations, we find an approximately 45\% maximum difference in the peak bolometric luminosity across mass models, a smaller spread than the order-of-magnitude differences found in single-$Y_e$ studies in the literature. However, this mass model variation induces a larger change in peak bolometric luminosity than variation in $Y_e$ in the lanthanide-rich regime ($Y_e \lesssim 0.30$), which includes our ensemble average $\langle Y_e\rangle = 0.247$.

\end{abstract}

\section{Introduction} 

Neutron star mergers are important astrophysical laboratories for probing some of the universe's most extreme conditions of matter. The inspiral and coalescence of the compact objects (binary neutron stars or a neutron star and a black hole) are detectable by ground-based gravitational wave observatories in the LIGO-Virgo-KAGRA (LVK) collaboration \citep{2020LVK}, and there is substantial evidence that compact binary mergers are the leading source of short-duration gamma-ray bursts (GRBs) \citep{Narayan_1992, Popham_1999, Fong_2013, Abbott_2017a}.

Compact mergers are also theorized to be a leading source of $r$-process nucleosynthesis \citep{Lattimer_1974, Freiburghaus_1999, Korobkin_2012, Metzger_2014}. The radioactive decay of $r$-process nuclei powers an electromagnetic transient called a kilonova, which provides an observational probe of the merger ejecta. Early models predict that neutron-rich ejecta produce an optical and infrared transient whose emission depends on the radioactive heating rate and the mass, velocity, and opacity of the ejecta \citep{Li_1998, Metzger_2010, Kasen_2013, Tanaka_2013}. AT2017gfo, following GW170817, is the first definitive kilonova observation, with data preferring multiple ejecta components including early blue emission followed by longer-lived red and infrared emission \citep{Cowperthwaite_2017, Kasen_2017, Smartt_2017, Villar_2017}.

The composition of the ejecta is determined by its entropy, expansion timescale, and electron fraction $Y_e = n_p/(n_p + n_n)$ (the number of protons per baryon). $Y_e$ sets the neutron richness of the outflow and hence the $r$-process abundance pattern \citep{Lippuner_2015}. Merger ejecta is not compositionally uniform, it arises from different physical channels with different thermodynamic histories. Dynamical ejecta launched during the merger can be neutron-rich enough to synthesize nuclei out to the second and third $r$-process peaks \citep{Korobkin_2012, Wanajo_2014, Radice_2018}, while neutrino irradiation in polar winds and post-merger disk outflows raises $Y_e$ and moves nucleosynthesis toward lighter $r$-process elements \citep{Metzger_2014, Perego_2014, Miller_2019}. Depending on their entropy and how strongly they are irradiated, however, disk outflows can still produce lanthanides and actinides \citep{Lippuner_2017}.
Simulations therefore predict a broad distribution of ejecta compositions rather than a single abundance pattern \citep{Radice_2018, Miller_2019, Lund_2024}. This distinction is important for the kilonova, since lanthanide- and actinide-rich material has significantly larger bound--bound opacity than lighter $r$-process material and produces redder, longer transients, while lanthanide-poor ejecta produces bluer, shorter emission \citep{Kasen_2013, Tanaka_2013, Metzger_2014, Kasen_2017}. Composition also sets the heating rate through the distribution of unstable nuclei \citep{Lippuner_2015}, so variations in the initial composition result in a diverse range of possible light curves and spectra \citep{Barnes_2021, Zhu_2021}.

The abundances and heating rates predicted by the $r$-process also depend on the nuclear physics inputs to the tool used to simulate it, particularly for nuclei far from stability where experimental constraints are limited \citep{DZ_1995, HFB27, HFB31, Moller_2016}. Nuclear mass models determine neutron separation energies and reaction $Q$-values (the energy released or absorbed in a reaction, given by the difference in rest mass energy between the initial and final states), which govern neutron capture, photodisintegration, and beta-decay pathways. Since different mass models predict slightly different shell closures, deformation properties, and drip line behavior, they can shift the abundance peaks and change the distribution of isotopes that power the kilonova at later times \citep{Moller_2016, Chen_2024}. Fission introduces further uncertainty in particularly neutron-rich ejecta, where the $r$-process flow reaches heavy nuclei that fission into lighter fragments. These fragments continue to capture neutrons and can be recycled back up to the fissioning region several times while free neutrons remain, so fission barriers and fragment distributions shape the final abundance pattern \citep{Goriely_2009, Mumpower_2018}. The resulting radioactive heating comes from beta decay, alpha decay, and fission: beta decay provides a strong source of heating throughout the kilonova \citep{Metzger_2010}, while alpha decay and fission become important at later times when heavier nuclei are prevalent, and hence matter more for lower initial $Y_e$ \citep{Mumpower_2018, Barnes_2021}. Not all of this decay energy is necessarily thermalized, since thermalization efficiency depends on the density and velocity of the ejecta (among other factors) \citep{Barnes_2016}. Changes in the initial composition or nuclear mass model can therefore alter both the total heating rate and which decay channel dominates at a given epoch.

Radioactive heating sets the energy source for the kilonova, but synthesizing light curves and spectra requires calculations that explicitly follow how ultraviolet to infrared photons diffuse through the rapidly expanding ejecta. Photon transport depends on the opacity, especially the bound--bound line opacity from heavy $r$-process elements, whose many overlapping lines create line ``forests.'' Since this opacity depends on composition, material with similar total heating rates can still produce different colors (i.e., different relative brightness across optical and infrared bands) and spectra \citep{Kasen_2013, Tanaka_2013}. Heating rates, light curves, and spectra must therefore be examined together to determine whether changes in kilonova observables result from radioactive heating, opacity effects, or both. Monte Carlo radiation transport methods are well suited to modeling kilonova ejecta, since synthesizing observable spectra does not require a dense, spatially global intensity solution \citep{Kasen_2013, Tanaka_2013, Wollaeger_2018}, and opacity regrouping and Monte Carlo diffusion techniques reduce the computational cost of line-dominated transport while retaining the major wavelength-dependent effects \citep{Cleveland_2014, Wollaeger_2014, Wagle_2023}.

Modern kilonova modeling couples nuclear reaction networks to simulate the $r$-process, which determine the time-dependent abundances and decay rates of isotopes \citep{Mumpower_2018, Sprouse_2021}, to radiation transport calculations, which use the resulting composition to compute opacities and determine how the decay power emerges as observable radiation \citep{Wollaeger_2013, Wollaeger_2014}. The networks rely on nuclear inputs such as masses, beta-decay properties, and fission barriers \citep{Moller_2003, Cyburt_2010, Moller_2015, Moller_2016}, many of which must be extrapolated for the neutron-rich nuclei most relevant to the $r$-process \citep{Kondev_2021, AME2020}. Uncertainties in both the ejecta composition and the nuclear physics therefore propagate into the same observables through the heating and the opacity \citep{Barnes_2021, Zhu_2021}. This overlap motivates a systematic study in which the two are varied independently. Varying both is especially important because $Y_e$ and nuclear physics can be degenerate in the observables they affect. For example, peak bolometric luminosity depends on the composition of the ejecta, which changes both with $Y_e$ and with the choice of nuclear mass model \citep{Zhu_2021, Barnes_2021}. Varying the two separately and comparing the results lets us test whether any observable breaks this degeneracy or if some degeneracy is broken in a specific $Y_e$ regime.

In this work, we investigate how variations in the initial ejecta composition and the nuclear mass model affect kilonova predictions during various observational epochs. We compute nucleosynthetic evolution and radioactive heating rates for a range of initial compositions and nuclear mass models, then propagate the resulting abundances into radiation transport calculations to obtain light curves and spectra. By comparing the heating rates with the resulting light curves and spectra, we separate changes caused primarily by radioactive power from those caused by opacity and transport effects. This comparison allows us to assess how astrophysical uncertainties in the ejecta composition and nuclear physics uncertainties in the mass model each contribute to the diversity of predicted kilonova emission.

The paper is organized as follows. In Sec.~\ref{sec:methods} we describe our setup, the software tools we use to simulate the $r$-process and perform radiative transfer, and the nuclear physics and composition variations. We discuss the effects of nuclear mass model variations on observables in Sec.~\ref{sec:nmmresults} and of composition variations in Sec.~\ref{sec:compresults}. We discuss our conclusions and future work in Sec.~\ref{sec:conclusions}.

\section{Methods} \label{sec:methods}

In the following subsections, we provide details of the methods and software used to build our kilonova model pipeline. Each subsection corresponds to a distinct step in the pipeline, as well as a distinct physical phenomenology or numerical method. Subsec.~\ref{subsec:tracers} provides the relevant information on the thermodynamic tracers we use; Subsec.~\ref{subsec: pl} describes how the tracers are mapped to a physically-motivated 1D spherical morphology; Subsec.~\ref{subsec: ns}, briefly summarizes the nucleosynthesis procedure for abundances; Subsec.~\ref{subsec: RT} provides the setup for the radiative transfer calculations for light curves and spectra; Subsec.~\ref{subsec: heating_rates} gives the details on computing radioactive decay heating rates and thermalization of the decay products.
Finally, Subsecs.~\ref{subsec:nmmvar} and~\ref{subsec:compvar} describe the approaches to nuclear mass model and composition variation we take in this study. Throughout the study, we refer to elements with $Z\geq100$ as superheavy elements, which is in contrast to similar studies such as \cite{Holmbeck_2023} which define superheavies as having $Z\geq104$.

\subsection{Tracers}\label{subsec:tracers}
\begin{figure}[h!]
    \centering
    \includegraphics[angle=0,width =1\linewidth]{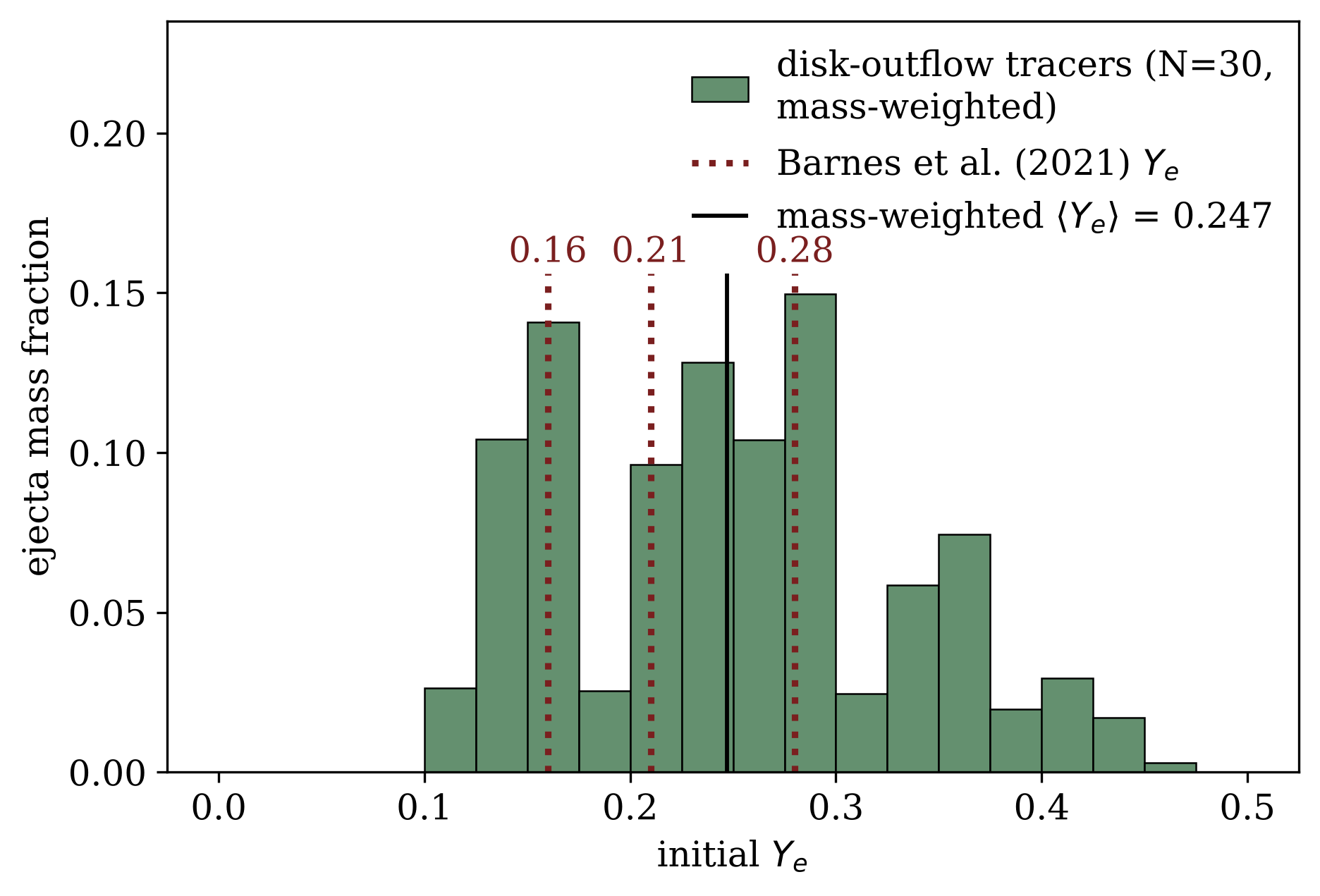}
    \caption{Histogram showing the mass-weighted $Y_e$ distribution for all 30 tracers, compared to the $Y_e$ values used in the \cite{Barnes_2021} study.}
    \label{yedistro}
\end{figure}

For studying the sensitivity of the kilonova observables to variations in the nuclear mass model, we use an established set of 30 tracers that represent post-merger disk conditions as computed and outlined in \cite{Lund_2024} and available at \cite{lund2025neutron}. The tracers are generated using the $\nu$\texttt{bhlight} general-relativistic neutrino radiation magnetohydrodynamics (GR$\nu$RMHD) code \citep{Miller_2019nu} with initial parameters representative of a remnant from GW170817 \citep{Abbott_2017a}, which includes an equilibrium torus of mass $M=0.12\: M_{\odot}$ with a constant $Y_e=0.1$ around a Kerr black hole of mass $M_{\rm BH}=2.58\:M_{\odot}$, and a dimensionless spin parameter of $a=0.69$ \citep{Miller_2019}. The torus is threaded with a single poloidal magnetic field
loop such that the minimum ratio of gas to magnetic
pressure is 10, indicative of a relatively strong magnetic field \citep{Lund_2024}. Each tracer represents a different radial shell of the ejecta cloud with its own mass, $Y_e$, velocity, temperature and density evolution. The tracers span velocities up to $0.2\: c$, and the total ejecta mass represented by all 30 tracers is $6.84\times10^{-3}\: M_{\odot}$. A histogram of the $Y_e$ distribution across the total mass is shown in Fig. \ref{yedistro} with a mass-weighted average $Y_e$ of 0.247. Although $\nu$\texttt{bhlight} models the post-merger disk and does not produce shock-heated dynamical ejecta, this $Y_e$ distribution spans a range of neutron-rich conditions similar to that found in the shock-heated and neutrino-processed ejecta of realistic NS-NS merger simulations \citep{Wanajo_2014, Radice_2018}. Each tracer is binned by \cite{lund2025neutron} into either a low velocity ($0-0.08\: c$) or high velocity ($0.08-0.2\: c$) bin, and otherwise does not have a resolved velocity or spatial coordinate. A description of the velocity distribution we use for radiative transfer is included in Subsec. \ref{subsec: pl}.

\label{grmhd}

\subsection{Two-Component Smoothly Broken Power-Law Density Profile}\label{subsec: pl}

Given that we do not have resolved space or velocity coordinates of the tracers, we must construct a physically motivated density profile in the velocity coordinate. To map the 30 angle-averaged tracers onto the homologously expanding ejecta required by the radiative transfer code we use (\texttt{SuperNu}, Subsec. \ref{subsec: RT}), we construct a 1D smoothly broken power law density profile. For details on the construction of this profile, see Appendix~\ref{subsec:A_powerlaw}. Critically for the comparison at the heart of this work, the tracer masses and bin memberships are identical across all seven nuclear mass models, so the solved break velocity, the radial shell grid, and the radial ordering are likewise identical for every model. Differences among the \texttt{SuperNu} calculations therefore arise exclusively from the microphysical inputs (the compositions, opacities, and heating rates) and not from the ejecta structure, which is held fixed by construction. For a comparison to similar studies which use a broken power-law construction, see \cite{Barnes_2013}, \cite{Kawaguchi_2020}, and \cite{Brethauer_2025}. Using this prescription allows us to construct a smooth physical model for the densities and velocities used for radiative transfer, while ensuring that the composition of each shell influences the observables in proper proportion to its GRMHD-derived mass in the kilonova evolution. 

\subsection{Nucleosynthesis}\label{subsec: ns}
To perform our nucleosynthesis calculations, we utilize the nuclear reaction network \texttt{PRISM} (Portable Routines for Integrated nucleoSynthesis Modeling) \citep{Mumpower_2017, Mumpower_2018, Sprouse_2021}. \texttt{PRISM} is designed to take nuclear data as well as the thermodynamic trajectory (temperature and density) and initial composition of a tracer to compute the relative abundances and mass fractions of nuclei as a function of proton number, atomic mass number, and time. We use the trajectories and initial compositions given in the tracer data from \cite{lund2025neutron}, which run $\nu$\texttt{bhlight} calculations out to 127 ms and extrapolate to 1000 s assuming homologous expansion, as our fiducial input to \texttt{PRISM}. We compute abundances for all tracers out to 1 Gyr. \texttt{PRISM} also outputs reaction flows for the main heating channels as they evolve over the course of the nucleosynthesis calculation: neutron capture, photodisintegration, beta decay, spontaneous fission decay, beta and neutron-induced fission decay, and alpha decay. We post-process this data to acquire heating rates as described in Subsec. \ref{subsec: heating_rates}. This heating rate data is used as a critical input to the radiative transfer simulation as described in Subsec. \ref{subsec: RT}.

\subsection{Radiative Transfer}
\label{subsec: RT}
In order to create spectra and light curves from our computed \texttt{PRISM} abundances, we use the Monte Carlo simulation code for simulating time-dependent non-linear radiative transfer interacting in matter, \texttt{SuperNu} \citep{Wollaeger_2013, Wollaeger_2014}. \texttt{SuperNu} is capable of taking multi-shell ejecta cloud data, opacity data, and source heating rates as input to simulate spectra and light curves. Our 1D input data consists of the velocity, mass, initial gas temperature, and nuclei abundances from \texttt{PRISM} for each tracer. Each tracer is assigned its velocity as discussed in Subsec. \ref{subsec: pl} in order to create a dataset in which the velocity is monotonically increasing as required by \texttt{SuperNu}. We give \texttt{SuperNu} realistic atomic opacities as tabulated in \cite{Fontes_2020} and \cite{Fontes_2023}. These tables are computed under the assumption of local thermodynamic equilibrium (LTE), and \texttt{SuperNu} correspondingly sets the thermal emissivity from the local gas temperature. To match the elements available in the opacity table, we reduce our \texttt{PRISM} abundance outputs by discarding elements with a mass fraction below $2\times10^{-7}$ and reassigning elements without tabulated opacities to chemically similar tabulated proxies in order to preserve total mass. The reassigned and discarded mass is negligible, $\lesssim 10^{-5}$ of each cell for all tracers. Since the GRMHD calculations from which we draw our tracers only run out to 127 ms, we take the temperature at that time and extrapolate it out to the starting time of our \texttt{SuperNu} simulations of 10000 seconds using an approximate adiabatic scaling given by $T=T_{127\: \text{ms}}\times(\frac{0.127\:\text{s}}{t})$, where $t$ is the starting time of the simulation in seconds. In testing, we find that by 10000 s the residual initial internal energy is a negligible fraction of the subsequently deposited radioactive energy, and the emergent light curves at observable epochs are insensitive to the choice of initial gas temperature, which is rapidly overwritten by the tabulated heating source. This tabulated heating source comes from \texttt{PRISM}, and includes the seven primary channels as discussed in Subsec. \ref{subsec: ns} (with the three fission channels summed into one heating rate). \texttt{SuperNu} does not internally update the elemental compositions as different isotopes decay. Rather, these decays enter through the \texttt{PRISM} heating rates. We initialize $2^{20}$ Monte Carlo particles, and radiation is transported over 1024 wavelength
groups spanning $0.1$--$12.8\,\mu$m ($10^{3}$--$1.28\times10^{5}$~\AA). The
emergent flux is tallied on a matching 1024-bin observer-frame wavelength grid assuming spherical symmetry. In convergence testing, we verify that the observables are unchanged to within Monte Carlo noise when the particle count is doubled.

\subsection{Heating Rates and Thermalization Fractions}
\label{subsec: heating_rates}
\texttt{PRISM} reports nuclide-resolved reaction flows rather than energy generation rates, so we compute channel-resolved specific heating in post-processing by weighting each flow with its reaction $Q$-value. The procedure is applied identically to every model and every tracer, so that heating differences between models are attributable to the mass surfaces and $Y_e$ variations alone. For each mass model we assemble a table of atomic masses $M(Z,A)$ in which the theoretical mass table populates the full grid and experimental masses from AME2020 \citep{AME2020} overwrite it wherever measurements exist, reproducing the theory-first, experiment-last assembly order used internally by \texttt{PRISM}.
For each reaction,
\begin{equation}
Q = \sum_{\rm reactants} M(Z,A) - \sum_{\rm products} M(Z,A),
\end{equation}
indexed by the heaviest reactant. Channels carrying a product yield
distribution (spontaneous and $\beta$-delayed fission) instead take the
yield-weighted product mass,
\begin{equation}
Q_{\rm fis} = M_{\rm parent} - \sum_{f} p_f\,M_f ,
\label{eq:qfis}
\end{equation}
so that $Q_{\rm fis}$ is the fragment-distribution-averaged fission $Q$-value. The specific heating rate in channel $c$ then follows from the flows
$F_{i,c}(t)$, tabulated in $\mathrm{mol\,g^{-1}\,s^{-1}}$, as
\begin{equation}
\dot{\epsilon}_c(t) = \mathcal{C}\sum_i Q_{i,c}\,F_{i,c}(t),
\label{eq:eps}
\end{equation}
where $\mathcal{C} = N_{\rm A}\,(1.602177\times10^{-6}\ \mathrm{erg\,MeV^{-1}})$,
with $\dot{\epsilon}_c$ in $\mathrm{erg\,g^{-1}\,s^{-1}}$ and the sum running
over all nuclides carrying flow at that timestep. The ensemble rate
supplied to the radiative transfer is the tracer mass-weighted mean,
\begin{equation}
\dot{\epsilon}_c^{\rm ens}(t) = \frac{\sum_j m_j\,\dot{\epsilon}_{c,j}(t)}
                                     {\sum_j m_j},
\end{equation}
evaluated on the union of the native tracer time grids with log--log
interpolation between tracer sample points.
\subsubsection{Neutrino losses and the \texorpdfstring{$\gamma$}{gamma}-ray channel}
Equation~(\ref{eq:eps}) returns the total nuclear energy release, which for
$\beta$ decay includes the antineutrino. Neutrinos free-stream out of the
ejecta and must be removed before the rates are used as a heating source.
We partition the $\beta$ channel as
\begin{equation}
\dot{\epsilon}_\beta \rightarrow
  \underbrace{f_{e}\,\dot{\epsilon}_\beta}_{\rm electrons} +
  \underbrace{f_{\gamma}\,\dot{\epsilon}_\beta}_{\rm decay\ \gamma} +
  \underbrace{f_{\nu}\,\dot{\epsilon}_\beta}_{\rm discarded},
\end{equation}
with $(f_{e}, f_{\gamma}, f_{\nu}) = (0.20,\,0.45,\,0.35)$ \citep{Barnes_2016, Mumpower_2018}. The neutrino share is discarded outright as neutrinos do not thermalize \citep{Barnes_2021}; the remainder is split between locally deposited $\beta$ electrons and decay $\gamma$ rays, which are transported rather than deposited in place. The $(n,\gamma)$ and $(\gamma,n)$ electromagnetic energies are routed alongside
the decay $\gamma$ rays. These are negligible for $t \gtrsim 0.1$~d and are fully reabsorbed at the optically thick earlier times, so the choice has little practical consequence.
\subsubsection{Density-dependent Thermalization Fractions}

\texttt{SuperNu} applies the \cite{Barnes_2016} thermalization fractions to the input heating rates detailed above, but does so in a per-cell  density dependent manner as given by the following formula:
\begin{equation}\label{thermfrac}
    f_{ch}=\frac{\ln(1+x)}{x}
\end{equation}
where x is the density-, thermalization coefficient-, and time-dependent variable

\begin{equation}\label{thermfracx}
    x=\frac{2b}{t_{center}*\rho_{cell}}.
\end{equation}

\begin{figure}
    \centering
    \includegraphics[width=1\linewidth]{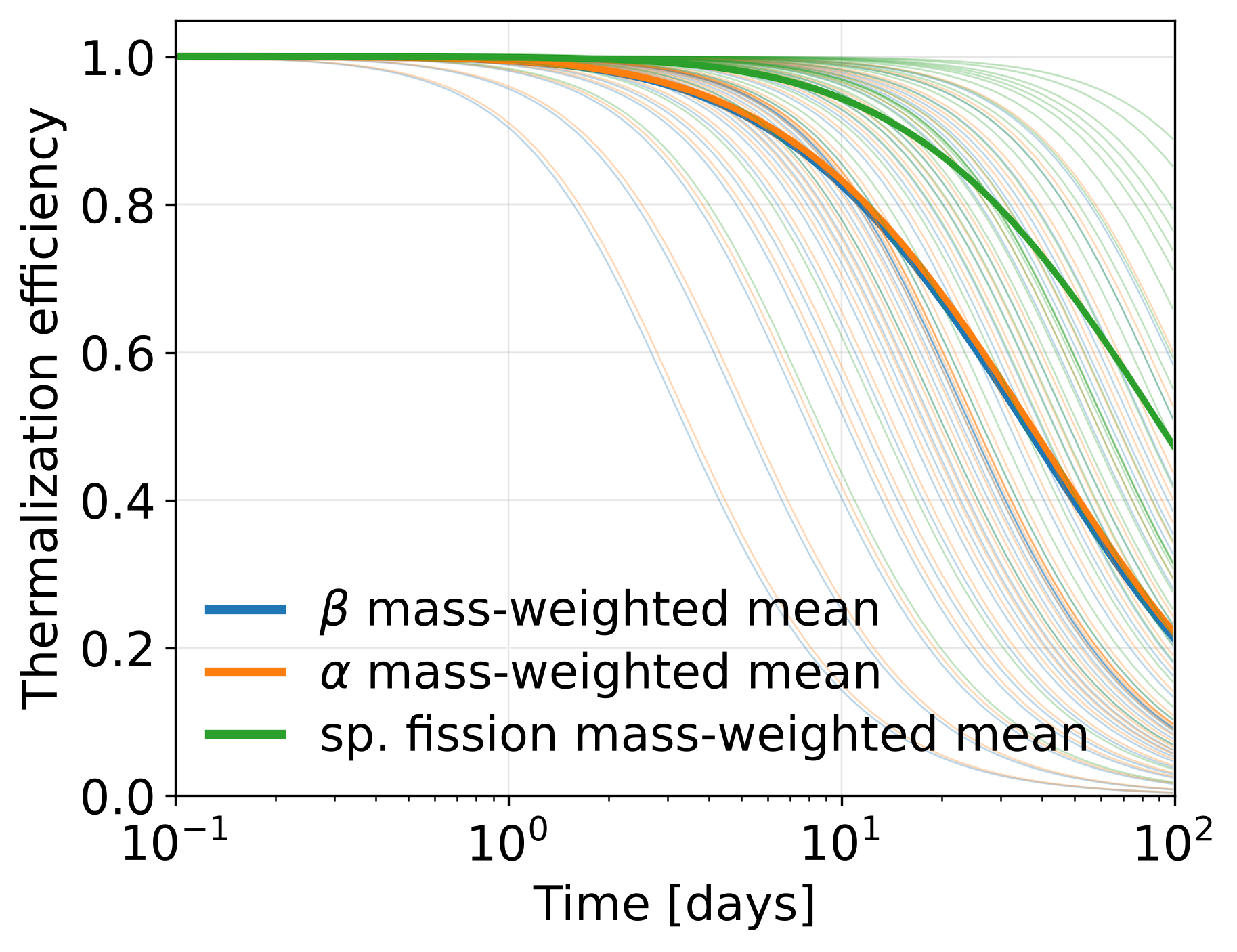}
    \caption{The per-decay channel thermalization efficiencies derived using Eqns. \ref{thermfrac} and \ref{thermfracx}. The darker lines show the mass-weighted mean curves, and the faint ones show the per-cell density-dependent thermalization fractions which \texttt{SuperNu} uses. Denser cells thermalize more efficiently.}
    \label{thermeff}
\end{figure}

Here, $b$ is a thermalization parameter that we select from \cite{Barnes_2016} for each decay channel: $b_{fission}=2.0\times10^{-12}$, $b_{\alpha}=1.2\times10^{-11}$, and $b_{\beta}=1.3\times10^{-11}$. In practice, this means that denser cells thermalize more efficiently.  We plot the mass-weighted and per-cell thermalization efficiency curves in Fig. \ref{thermeff}, which shows that spontaneous fission is considerably more efficient at heating the ejecta than beta and alpha decay. Alpha decay is only slightly more efficient than beta decay using this model with these parameters.

\subsection{Nuclear Mass Models} \label{subsec:nmmvar}
The choice of nuclear mass model affects $r$-process nucleosynthesis simulation outputs as well as heating rates in the kilonova ejecta cloud \citep{Mumpower_2015, Barnes_2021, Zhu_2021, Chen_2024}. In this study, we use the seven theoretical nuclear mass models and fission barrier prescriptions given in Tab. \ref{nucmods}. All of these mass models are used as nuclear data input to \texttt{PRISM}, and are used to compute their corresponding heating rates to be used as source input to \texttt{SuperNu} as mentioned in Subsec. \ref{subsec: RT}. We use data from the JINA Reaclib library \citep{Cyburt_2010} for various channel reactions (e.g., ($\alpha$,~$\gamma$), ($\gamma$,~$\alpha$), ($n$,~$\gamma$), etc.).  Where possible, we overwrite theoretical data with experimental data from the 2020 NuBase dataset \citep{Kondev_2021}. In particular, we use $\beta-$ decays with delayed neutron probabilities from \cite{Moller_2003} in conjunction with NuBase $\beta-$ decay rate data. We also use supplemental NuBase data for spontaneous fission reactions, $\beta+$ decays, electron capture rates, spontaneous neutron and proton emission rates, and alpha decay rates. Experimental masses from the AME2020 evaluation \citep{AME2020} are used, where available, in computing the neutron capture and photodisintegration rates. We use theoretical spontaneous fission reaction data from \cite{Karpov_2012} with finite-range liquid-drop model (FRLDM) fission barriers only for nuclei for which NuBase does not contain measured half-lives. The Reaclib and NuBase contributions to the nuclear datasets are consistent for every nuclear mass model variation. Now for each model, we compute theoretical $\beta-$ decays, neutron capture and photodisintegration rates, $\beta$-delayed fission reactions, neutron-induced fission reactions, and Viola-Seaborg alpha decay rates, all calculated consistently using each model's respective masses. We compute the neutron-induced fission and neutron capture rates using the statistical Hauser-Feshbach software CoH \citep{Kawano_2016}, and the $\beta-$ decays and $\beta$-delayed fission reactions using the companion code BeoH \citep{Lovell_2021}.

\startlongtable
\begin{deluxetable}{l c r}
\tablewidth{0pt}
\tablecaption{\label{nucmods}Mass Models Used in the Nucleosynthesis Simulations and Heating Rate Calculations.}
\tablehead{\colhead{Mass Model} &  \colhead{Abbreviation} & \colhead{Masses, Fission Barriers}}
\startdata
Finite-Range Droplet Model & FRDM2012 & ~\citet{Moller_2016,Moller_2015}\\
Duflo and Zuker 33 & DZ33 & ~\citet{DZ_1995, Moller_2015}\\
Hartree-Fock-Bogoliubov 27 & HFB27 & \citet{HFB27, Goriely_2009}\\
Hartree-Fock-Bogoliubov 31 & HFB31 & \citet{HFB31, Goriely_2009}\\
Skyrme-HFB with UNEDF1 & UNEDF1 & \citet{UNEDF1, Moller_2015}\\
Skyrme-HFB with SLy4 & SLy4 &\citet{SLy4_1, Moller_2015}\\
Weizs\"{a}cker-Skyrme 4 &WS4& \citet{WS4, Moller_2015}\\
\enddata
\end{deluxetable}
\twocolumngrid

\subsection{Composition Variations with Fixed Ejecta Structure}
\label{subsec:compvar}

In contrast to the fixed composition used for the nuclear mass model portion of this study, here we vary only the composition assigned to each velocity component and its associated radioactive heating, while keeping the total ejecta mass, the mass of each velocity component, and the density profile (Subsec.~\ref{subsec: pl}) fixed. This isolates the effect of the ejecta composition from changes in the ejecta density and velocity structure. Throughout our composition variation study, we use the FRDM2012 nuclear mass model.

Following the two radial-velocity bins of Subsec.~\ref{subsec: pl}, we identify
two velocity components: the slow velocity component $\mathcal{S}$
($v < 0.08c$) and the fast velocity component $\mathcal{F}$ ($0.08c < v < 0.2c$). The density contribution of each velocity component is held fixed throughout the composition grid, so the velocity-component masses
\begin{equation}
M_{\mathcal{S}} = \int \rho_{\mathcal{S}}(\mathbf{x})\,d^3x, \qquad
M_{\mathcal{F}} = \int \rho_{\mathcal{F}}(\mathbf{x})\,d^3x,
\label{eq:component_masses}
\end{equation}
and the total ejecta mass $M_{\rm ej}=M_{\mathcal{S}}+M_{\mathcal{F}}$ are the same in every run.

The tracer set of \citet{Lund_2024} comprises 30 tracers: 15 $Y_e$
bins, each sampled in both velocity components. We label each tracer by the
pair $(i,c)$, where $i=1,\ldots,15$ indexes the $Y_e$ bin and
$c\in\{\mathcal{S},\mathcal{F}\}$ indexes the velocity component. The
\texttt{PRISM} calculation for each tracer is performed once and reused without modification for every composition variation in this section (i.e., no new nucleosynthesis is computed), giving a single final abundance pattern $X_{i,c}(Z,A)$ and
radioactive heating rate $\dot{q}_{i,c}(t)$ for every tracer; only the weights with which the tracers are combined change.

For the fiducial composition of each velocity component we take a mass-weighted average over the tracers belonging to that velocity component, writing $m^0_{i,c}$
for the fiducial ejecta mass associated with tracer $(i,c)$ so that
$\sum_{i=1}^{15} m^0_{i,c} = M_c$ (see Appendix~\ref{subsec:A_fiducial}
for the explicit expressions).
We construct the composition variations by reassigning the tracer weights that define the composition of each velocity component while conserving the total mass of that velocity component. Let $\mu_{i,c}$ denote the composition weight, in units
of mass, assigned to tracer $(i,c)$. The weights satisfy
\begin{equation}
\sum_{i=1}^{15} \mu_{i,\mathcal{S}} = M_{\mathcal{S}}, \qquad
\sum_{i=1}^{15} \mu_{i,\mathcal{F}} = M_{\mathcal{F}},
\label{eq:weight_norm}
\end{equation}
so that the varied compositions and heating rates,
\begin{equation}
\bar{X}_{c}(Z,A) = \frac{1}{M_{c}} \sum_{i=1}^{15}
\mu_{i,c}\, X_{i,c}(Z,A),
\label{eq:varied_comp}
\end{equation}
\begin{equation}
\bar{\dot{q}}_{c}(t) = \frac{1}{M_{c}} \sum_{i=1}^{15}
\mu_{i,c}\, \dot{q}_{i,c}(t),
\label{eq:varied_heat}
\end{equation}
remain properly normalized weighted averages. Using the same weights $\mu_{i,c}$ in both
Eqs.~(\ref{eq:varied_comp}) and (\ref{eq:varied_heat}) ensures that the
composition and the radioactive heating of each velocity component stay mutually
consistent. Setting $\mu_{i,c}=m^0_{i,c}$ recovers the fiducial case of
Eqs.~(\ref{eq:fid_comp}) and (\ref{eq:fid_heat}).

To build the grid we adjust the relative weights of the tracers in each velocity component to obtain the target mass-weighted mean $Y_e$. 

\begin{figure}[tb]
    \centering
    \includegraphics[width=1\linewidth]{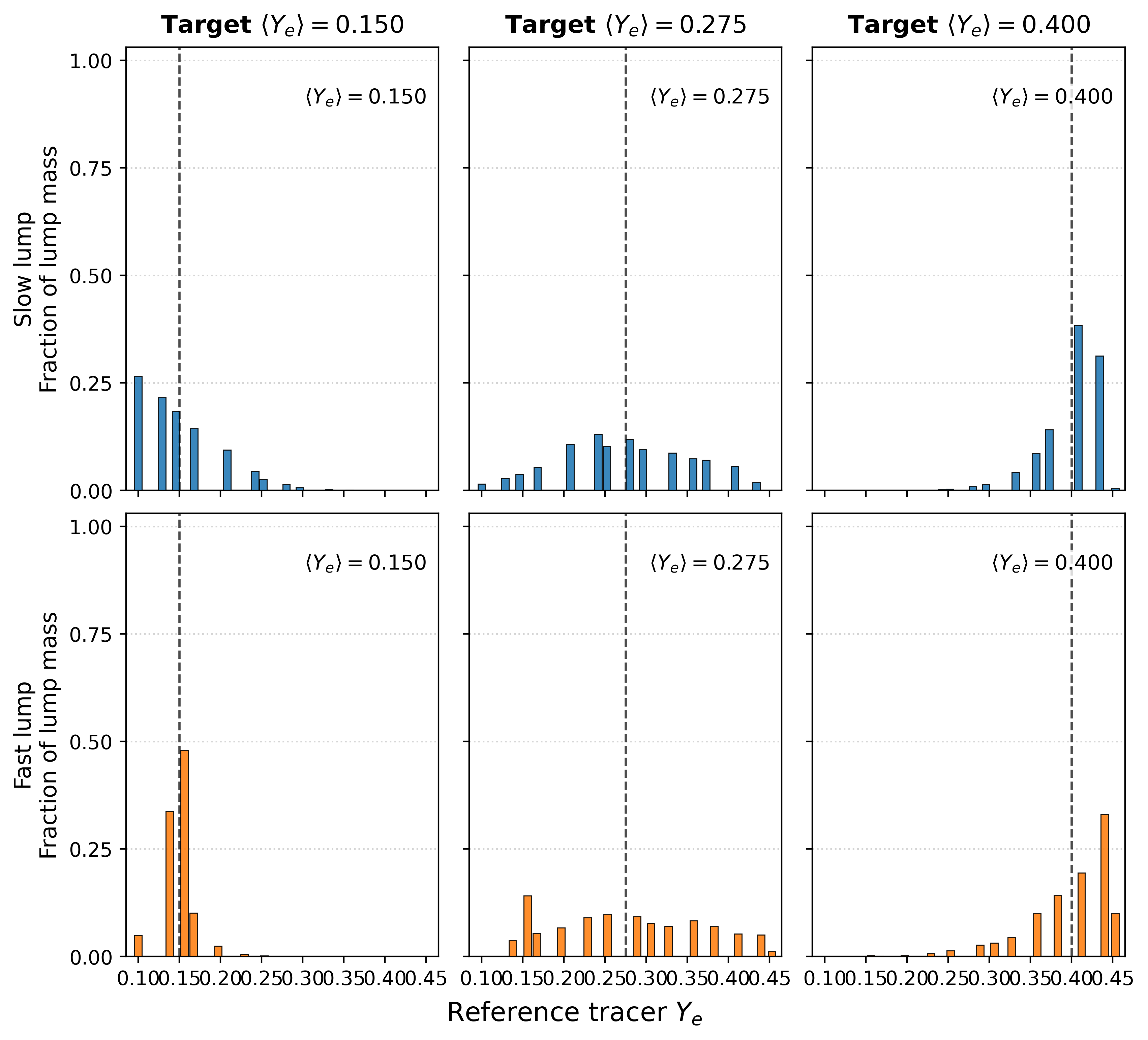}
    \caption{Tracer composition weights $\mu_{i,c}$ in the slow (top row) and
    fast (bottom row) velocity components for three target $Y_e$ values.
    Increasing $\beta_c$ shifts weight toward tracers with higher $Y_e$ while conserving the velocity component mass.}
    \label{fig:compweights}
\end{figure}

Using the reference electron fraction $Y^{\rm ref}_{e, i , c}$ of each tracer, we define
\begin{equation}
    p_{i, c} (\beta_c) = \frac{m^0_{i, c} \exp(\beta_c Y^{\rm ref}_{e, i, c})}{\sum_{j=1}^{15} m^0_{j, c} \exp(\beta_c Y^{\rm ref}_{e,j,c})}
\end{equation}
and
\begin{equation}
    \mu_{i, c} = M_c \left[\epsilon + (1-15\epsilon) p_{i, c} \right],
\end{equation}
where here we take $\epsilon = 10^{-12}$. This floor retains each tracer as a positive weight. For each electron fraction tested, we choose $\beta_c$ separately for the slow and fast components to ensure $M_c^{-1}\sum_i\mu_{i,c}Y^{\rm ref}_{e,i,c}$ meets the desired electron fraction. We then account for the floor contribution and solve for $\beta_c$. The fiducial model naturally retains the original weighting from the tracers. 
Example weight distributions for three target $Y_e$ values are shown in Fig.~\ref{fig:compweights}. 

The composition assigned to a spatial cell follows from the density
contributions of the two velocity components. Where both velocity components contribute
to a cell, the cell composition is the density-weighted average
\begin{equation}
X_{\rm cell}(Z,A)=
\frac{\rho_{\mathcal{S}}(\mathbf{x})\bar{X}_{\mathcal{S}}(Z,A)
     +\rho_{\mathcal{F}}(\mathbf{x})\bar{X}_{\mathcal{F}}(Z,A)}
     {\rho_{\mathcal{S}}(\mathbf{x})+\rho_{\mathcal{F}}(\mathbf{x})},
\label{eq:cell_composition}
\end{equation}
and the heating rate is weighted identically,
\begin{equation}
\dot{q}_{\rm cell}(t)=
\frac{\rho_{\mathcal{S}}(\mathbf{x})\bar{\dot{q}}_{\mathcal{S}}(t)
     +\rho_{\mathcal{F}}(\mathbf{x})\bar{\dot{q}}_{\mathcal{F}}(t)}
     {\rho_{\mathcal{S}}(\mathbf{x})+\rho_{\mathcal{F}}(\mathbf{x})}.
\label{eq:cell_heating}
\end{equation}

This procedure preserves the total ejecta mass, the slow and fast velocity-component
masses, and the density profile, and therefore isolates the effect of changing
the initial composition and the associated heating. The cost of the
simplification is that spatial variation of the composition within a
velocity component is lost: the fiducial point of this grid corresponds to a
velocity-component-averaged version of the original tracer mapping rather than the full 30-tracer map.

\section{Results: Nuclear Mass Model Variations}\label{sec:nmmresults}
We now discuss the results from the nuclear mass model variation portion of our study. Specifically, we examine the mass model effect on abundances (Subsec.~\ref{subsec:nmmyields}), heating rates (Subsecs.~\ref{subsec:nmmhrv}-\ref{subsec:alfnuc}), and light curves and spectra (Subsecs.~\ref{subsec:nmmvlc}-\ref{subsec:nmmvspec}), respectively. Hereafter we denote the lanthanide and actinide mass fractions by $X_{\rm lan}$ and $X_{\rm act}$, and the mass fraction of elements with $Z\geq100$ by $X_{Z\geq100}$.

\begin{figure}[h!]
    \centering
    \includegraphics[angle=0,width =1\linewidth]{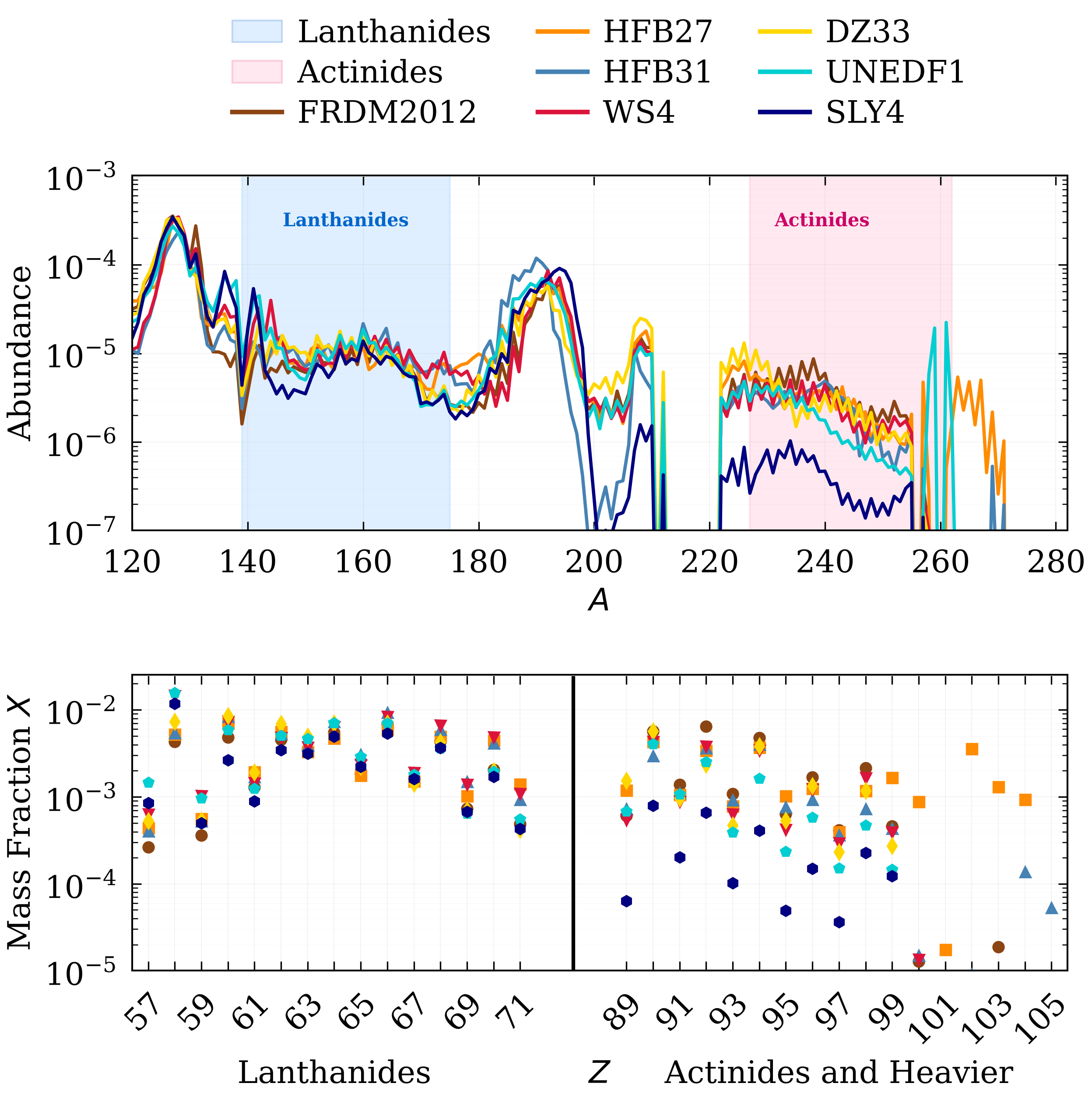}
    \caption{Nucleosynthesis yields between all nuclear mass models used in this study. The top panel shows the mass-weighted abundance curves for each mass model, with the lanthanide and actinide atomic mass number ranges highlighted. The bottom panel shows the mass-weighted mass fractions of each element within the lanthanides and actinides, and some heavier elements up to $Z=105$. There is no model which clearly produces the most lanthanides nor actinides, but HFB27 produces by far the most superheavy elements ($Z\geq100$).}
    \label{fig: abundances}
\end{figure}

\subsection{Nuclear Mass Model Nucleosynthesis Yields}\label{subsec:nmmyields}

\begin{figure*}[t]
    \centering
    \begin{subfigure}[t]{0.48\textwidth}
        \centering
        \includegraphics[width=\textwidth]{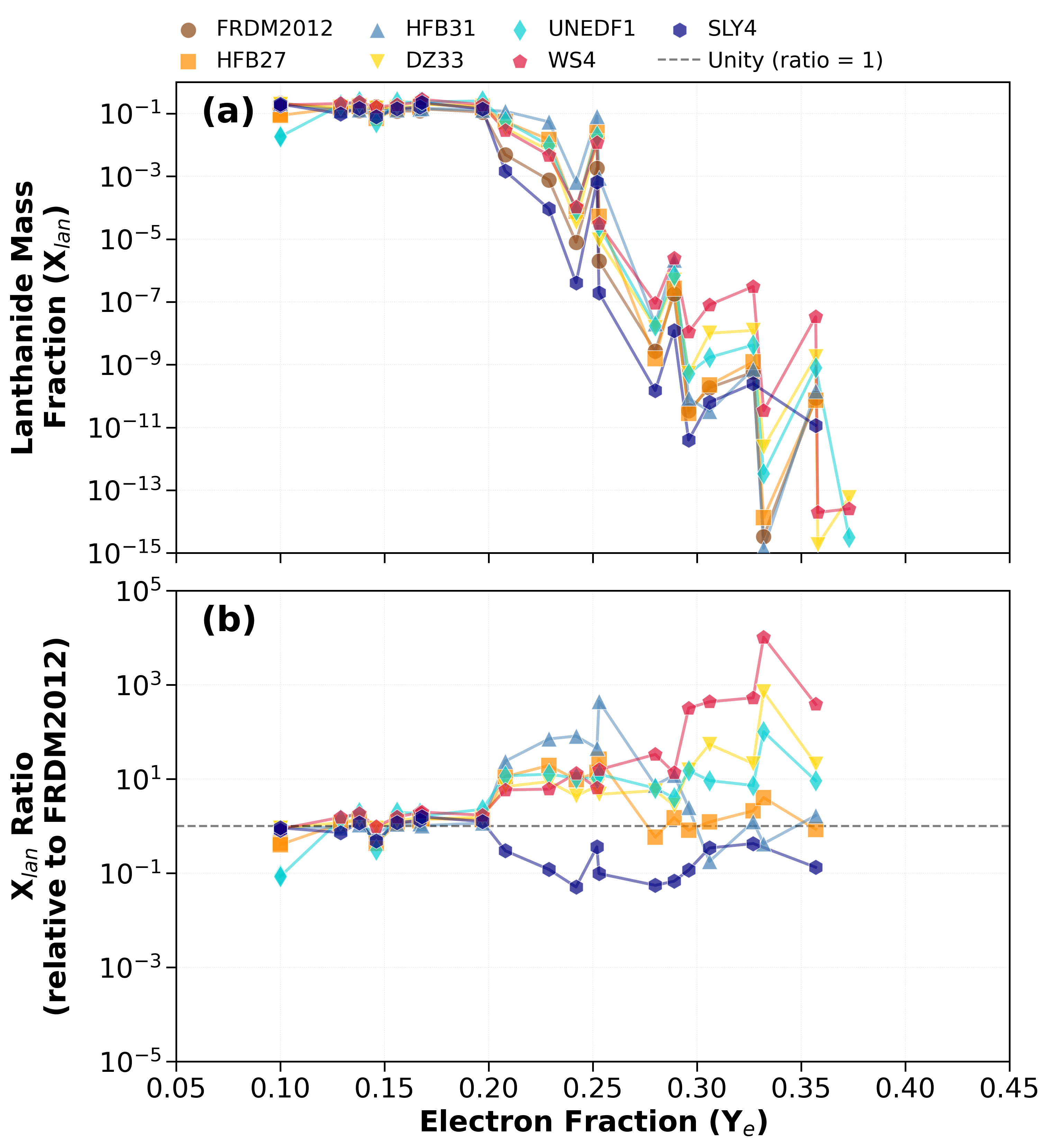}
        \caption{Lanthanide mass fractions}
        \label{fig:lanthanides}
    \end{subfigure}
    \hfill
    \begin{subfigure}[t]{0.48\textwidth}
        \centering
        \includegraphics[width=\textwidth]{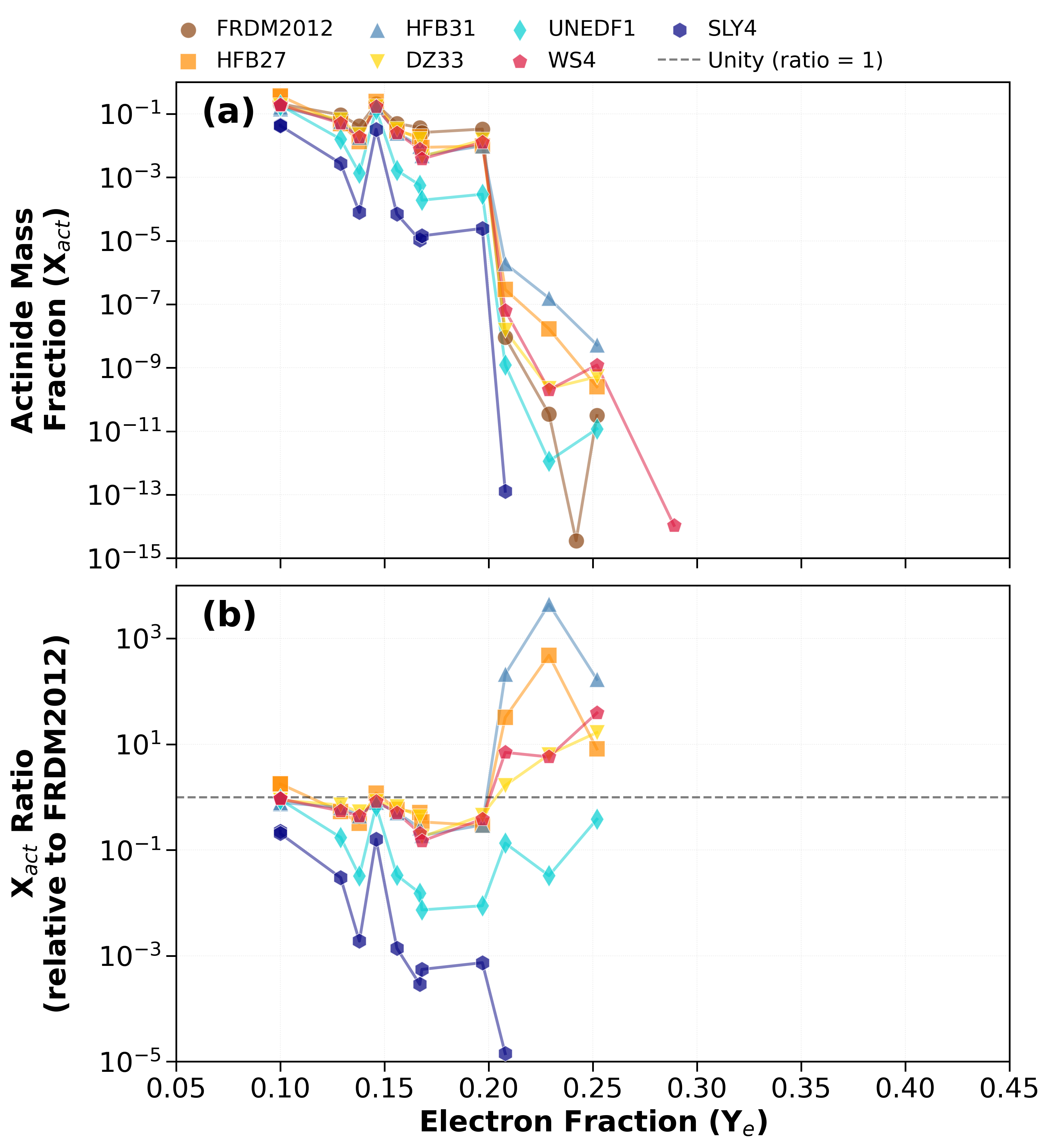}
        \caption{Actinide mass fractions}
        \label{fig:actinides}
    \end{subfigure}
    \caption{Mass fractions and ratios as a function of electron fraction (Y$_e$) for lanthanides and actinides. 
    (a) Lanthanides: Upper panel shows X$_{\rm lan}$ vs Y$_e$, lower panel shows ratios relative to FRDM2012.
    (b) Actinides: Upper panel shows X$_{\rm act}$ vs Y$_e$, lower panel shows ratios relative to FRDM2012.}
    \label{fig:lanthact}
\end{figure*}

\begin{figure}
    \centering
    \includegraphics[angle=0,width =1\linewidth]{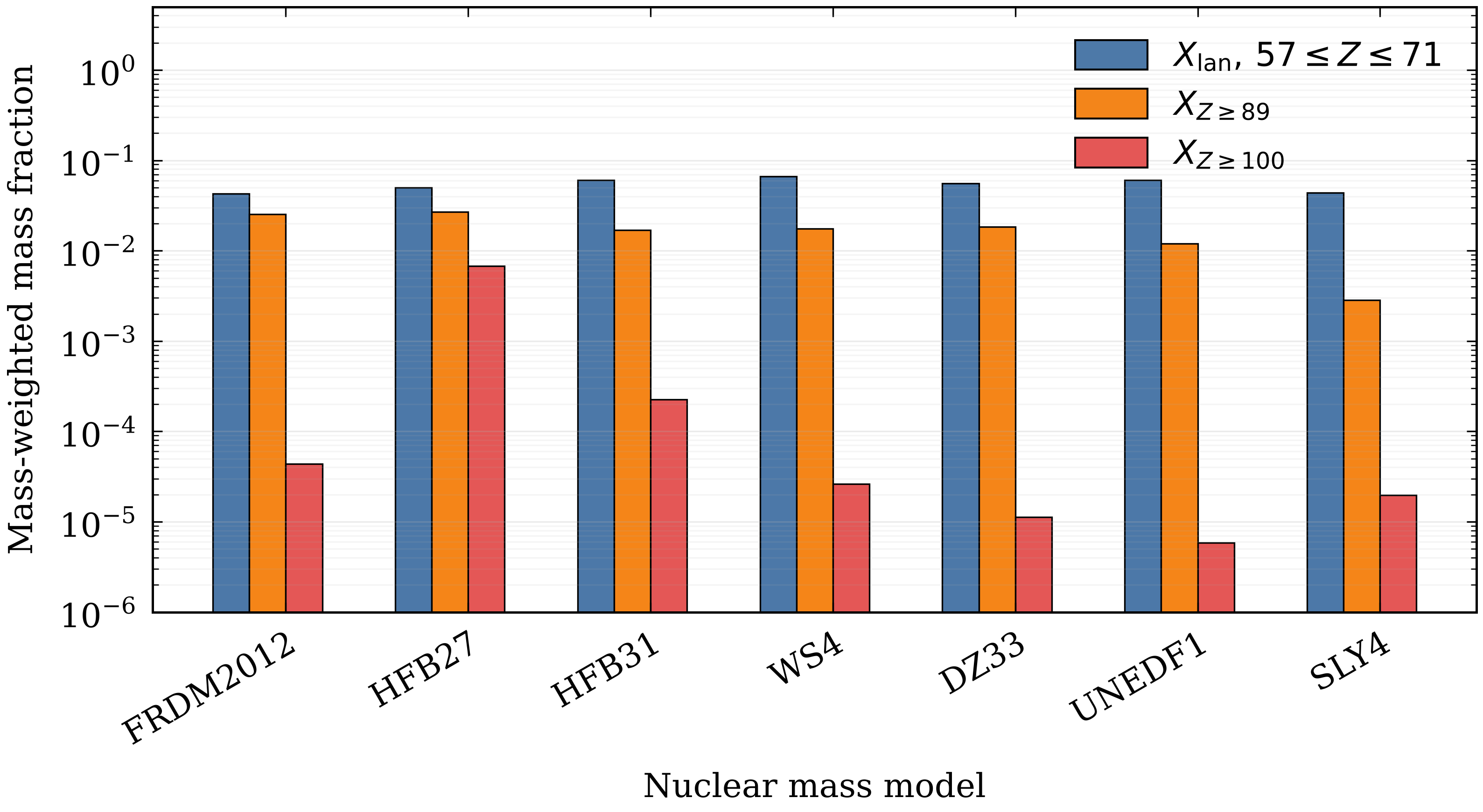}
    \caption{Histogram showing the mass-weighted mass fractions produced by each mass model, binned by elements with $57\leq Z\leq71$ (lanthanides, blue), $Z\geq89$ (orange), and $Z\geq100$ (red). There is visually almost no difference in the lanthanide production between mass models, little noticeable difference for $Z\geq89$, and significant differences for $Z\geq100$.}
    \label{fig: xlanxactmw}
\end{figure}

In this section, we discuss the production of elements from our \texttt{PRISM} simulations for the nuclear mass models shown in Tab. \ref{nucmods}. Fig. \ref{fig: abundances} shows the mass-weighted abundance curves for each nuclear mass model as well as the lanthanide and actinide mass fractions at their respective $Z$ for every model at 1 day. At this time, there are still a significant number of spontaneously fissioning elements with relatively short half-lives (i.e. $Z\geq94$) present in the ejecta. Extracting data at such an epoch is important since it allows us to see how strongly the spontaneous fission of heavier elements affects the observables for each mass model. Figure~\ref{fig:lanthact} shows the lanthanide and actinide fractions as functions of $Y_e$. Figs.~\ref{fig: abundances} and \ref{fig:lanthact} show that the SLy4 model has the lowest mass fraction of both lanthanides and actinides across the studied set. 
HFB27 produces the most elements with $98<Z\leq104$, and produces a significant mass fraction of elements with $Z\geq100$ alongside HFB31 which produces a significant ($\sim10^{-4}$) mass fraction of Rf and Db. Other general trends across mass models are broadly consistent with \cite{Zhu_2021}, \cite{Barnes_2021}, and \cite{Chen_2024}. The differences in $X_{\rm lan}$ between models are much smaller for the mass-weighted ensemble (Fig.~\ref{fig: abundances}) than for individual tracers (Fig.~\ref{fig:lanthact}), so model-to-model variations in $X_{\rm lan}$ largely cancel in the tracer ensemble. $X_{\rm act}$ follows a similar trend, but with an important distinction from $X_{\rm lan}$. As can be seen in Figs. \ref{fig: abundances} and \ref{fig:lanthact}, no single model dominates in the production of lanthanides or actinides across the entire $Y_e$ distribution. However, as HFB27 produces by far the largest mass fraction of elements with $Z\geq100$, a closer look at the actinide production behavior is warranted. Plotting the mass-weighted mass fraction for each model (Fig. \ref{fig: xlanxactmw}) shows that HFB27 has a significant mass excess in comparison to the other models for $Z\geq100$. This production of such heavy and unstable isotopes is expected to produce a much higher spontaneous fission (SPF) heating rate for HFB27 across time, and particularly the excess of elements which are long-lived fissioning isotopes such as ${}^{254}_{98}\mathrm{Cf}$ and ${}^{271}_{104}\mathrm{Rf}$ which have been found to significantly impact the spontaneous fission heating rate after $t\approx10$ days \citep{Barnes_2021, Zhu_2021, Zhu_2018}. Thus, we expect this excess of heavy elements to create significant difference in the spontaneous fission heating rates between mass models as explored in Subsecs. \ref{subsec:nmmhrv} and \ref{subsec:spfnuc}.

\begin{figure*}[t]
    \centering

    \begin{subfigure}{0.48\textwidth}
        \centering
        \includegraphics[width=\linewidth]{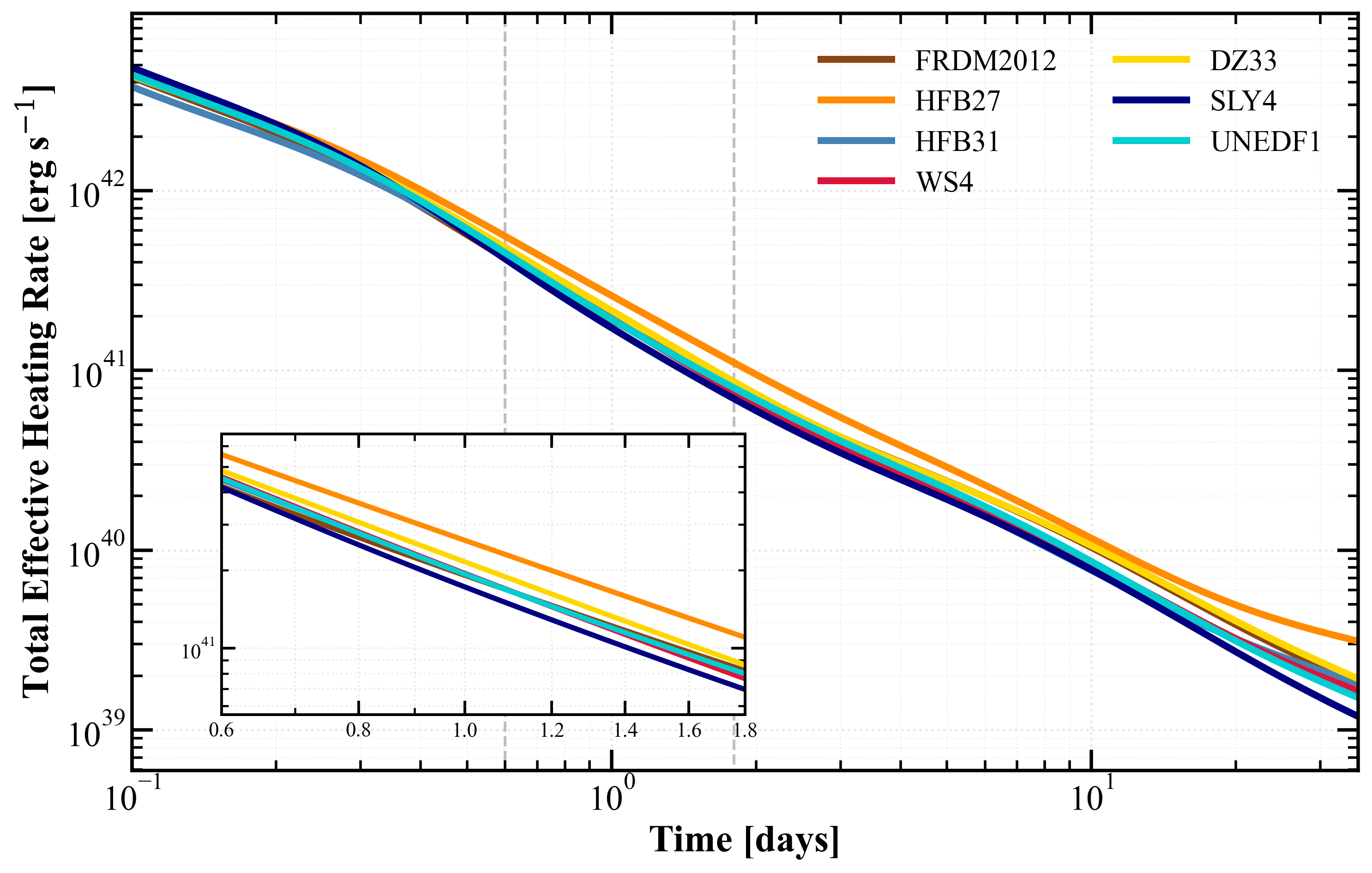}
        \caption{Total effective heating rate.}
        \label{fig:total-eff-heating}
    \end{subfigure}
    \hfill
    \begin{subfigure}{0.48\textwidth}
        \centering
        \includegraphics[width=\linewidth]{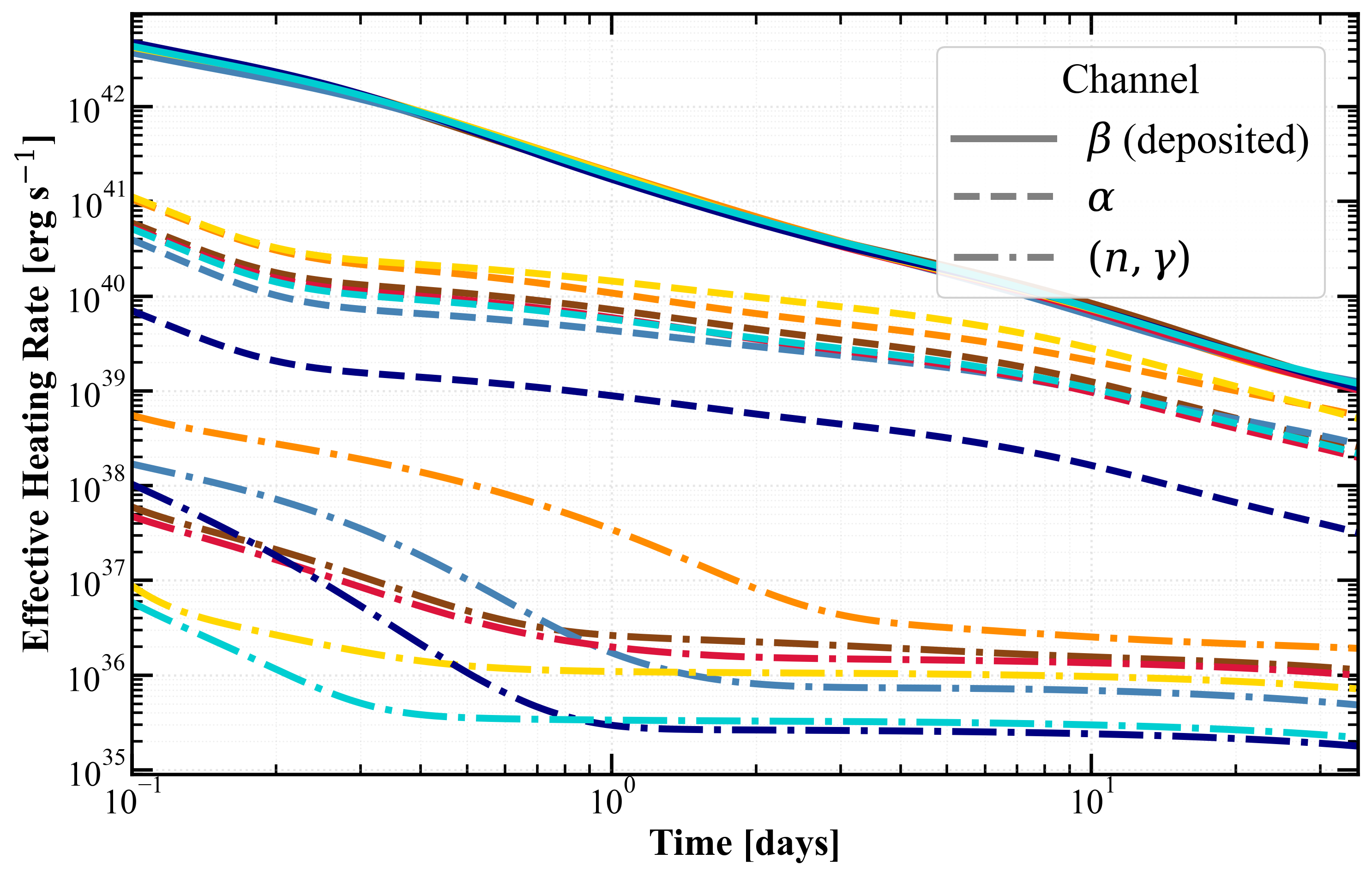}
        \caption{Per-channel effective heating rate.}
        \label{fig:channel-heating}
    \end{subfigure}

    \caption{Mass-weighted effective heating rates across different mass models. (a) Total heating rate summed over all channels. (b) Individual contributions from beta decay (solid), alpha decay (dashed), and neutron capture (dash-dot). Mass models do not diverge significantly in their total effective heating rate, but do when broken down into individual channels.}
    \label{fig:heating-rates}
\end{figure*}

\begin{figure*}
    \centering
    \includegraphics[width=1\linewidth]{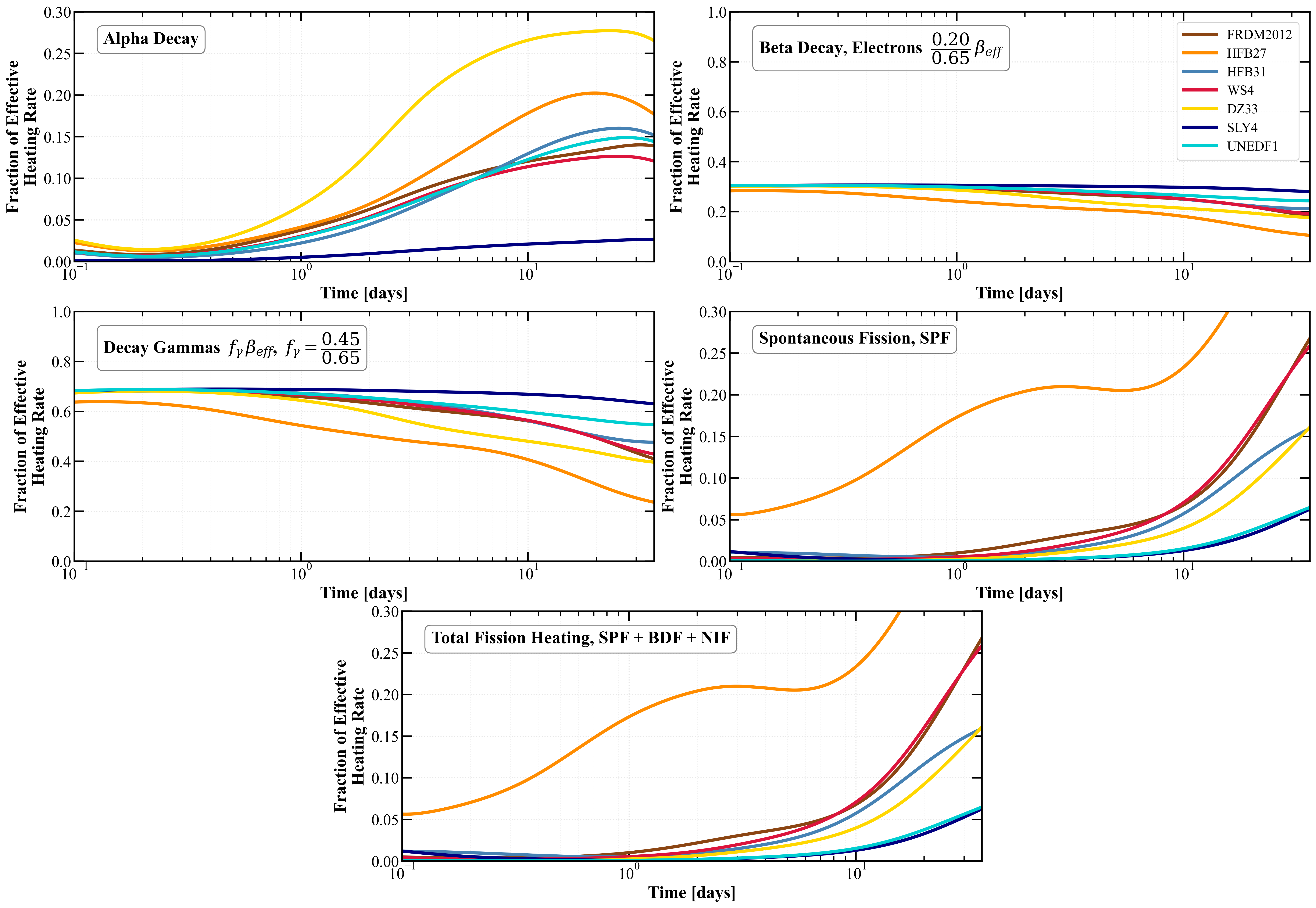}
    \caption{Fractional heating rates for the alpha (top left), beta (electrons, top right), beta (gamma, middle left), spontaneous fission (middle right), and total fission heating (spontaneous, beta-delayed, and neutron induced fission, bottom) channels. HFB27 is a significant outlier in fission heating, and also in alpha decay heating. DZ33 has the highest alpha decay heating rate fraction, and SLy4 has the lowest. UNEDF1 and SLy4 have the smallest fission heating fractions.}
    \label{fig:frac-heating}
\end{figure*}

\begin{figure}
    \centering
    \includegraphics[width=1\linewidth]{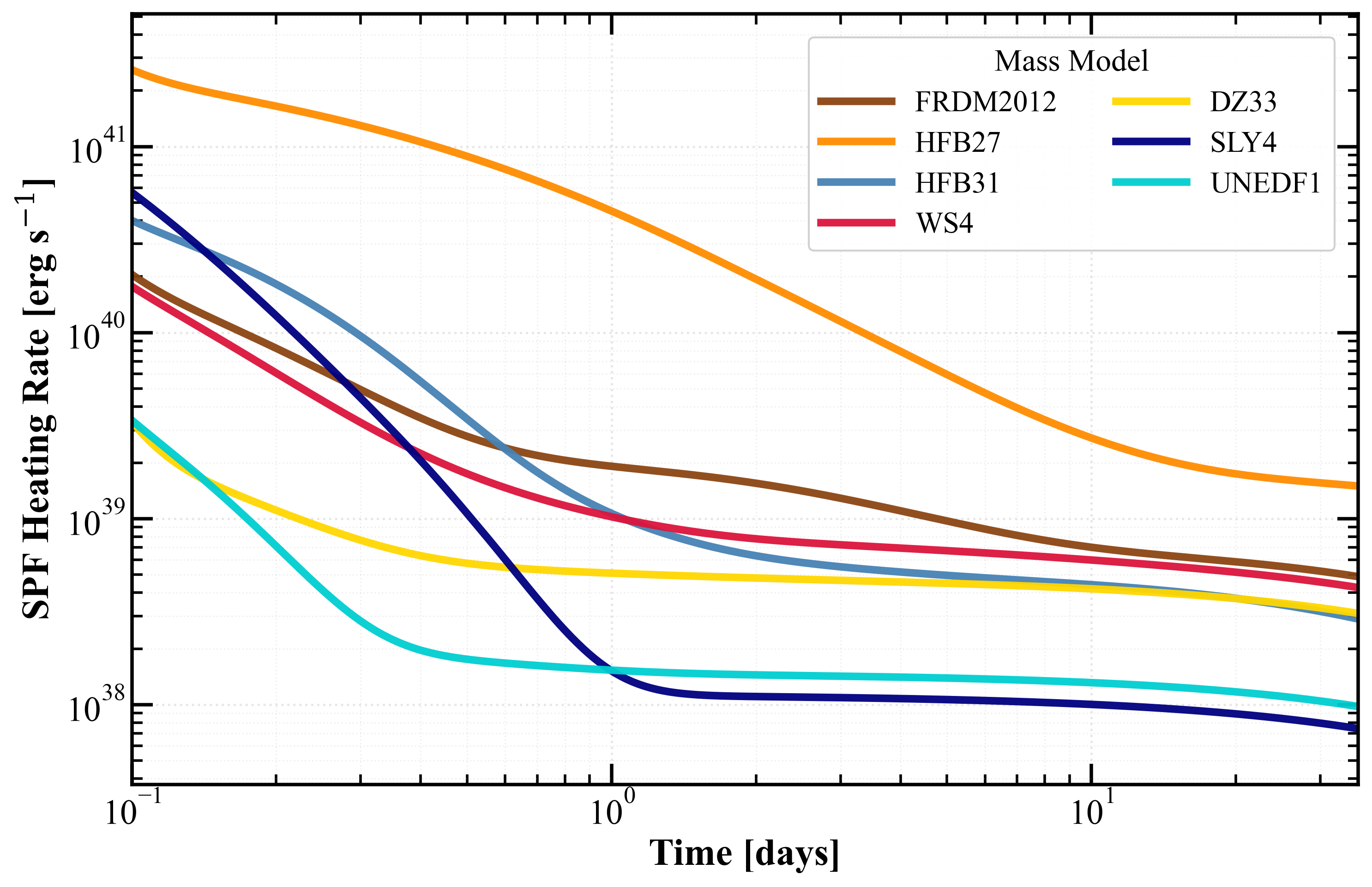}
    \caption{The spontaneous fission heating rate for each mass model. HFB27 has the largest SPF heating by a large margin, and SLy4 and UNEDF1 have the lowest as also shown in Fig. \ref{fig:frac-heating}.}
    \label{fig:spfheating}
\end{figure}

\subsection{Nuclear Mass Model Heating Rate Variations}\label{subsec:nmmhrv}
Differences in the elemental abundances produced across mass models are expected to have a large impact on the effective heating rate of an ejecta cloud as detailed in \cite{Zhu_2021} and \cite{Barnes_2021}. This is in large part due to the spontaneous fissioning of heavy elements, which some models (such as HFB27) produce a significant amount of as discussed in Subsec. \ref{subsec:nmmyields}. Fig. \ref{fig:total-eff-heating} shows the total effective heating rate for each model, mass-weighted across all 30 tracers. These total effective heating rates are similar across the time scale of interest (35 days) with no clear outlier, until later times when HFB27 begins to diverge. Looking at the breakdown of effective heating rates across channels (Figs. \ref{fig:channel-heating} and \ref{fig:frac-heating}) and the heating rate due to fission in particular (Fig. \ref{fig:spfheating}) show a clear divergence among mass models in all heating rates. We also find that $\beta$ decay heating is dominant in value across the entire timescale of interest, consistent with the findings of \cite{Zhu_2021}. As mentioned in Subsec. \ref{subsec:nmmyields}, the HFB27 model produces significantly more of the spontaneously fissioning elements and is the only model to produce a large amount of elements with $Z\geq100$ whose primary decay mode is spontaneous fission. This drives the large, order of magnitude difference in the spontaneous fission heating rate between HFB27 and the other nuclear mass models. We also note that on the opposite end of the spectrum, SLy4 and UNEDF1, which produce the lowest abundances of fissioning actinides relative to the other models, correspondingly have the lowest fission heating rate. We also plot the fractional heating rates for each channel in Fig. \ref{fig:frac-heating} to show the increasing dominance of fission and $\alpha$ heating as time progresses, in line with the findings of \cite{Barnes_2021}. The main takeaway from these results is that there are small differences in the tracer mass-weighted total effective heating rates for every nuclear model, but larger ones in the $\alpha$ and SPF fission heating rates which could drive significant differences in the observable signatures of the mass models in this realistic $Y_e$ distribution ejecta cloud.

\begin{figure}[h!]
    \centering
    \includegraphics[angle=0,width =1\linewidth]{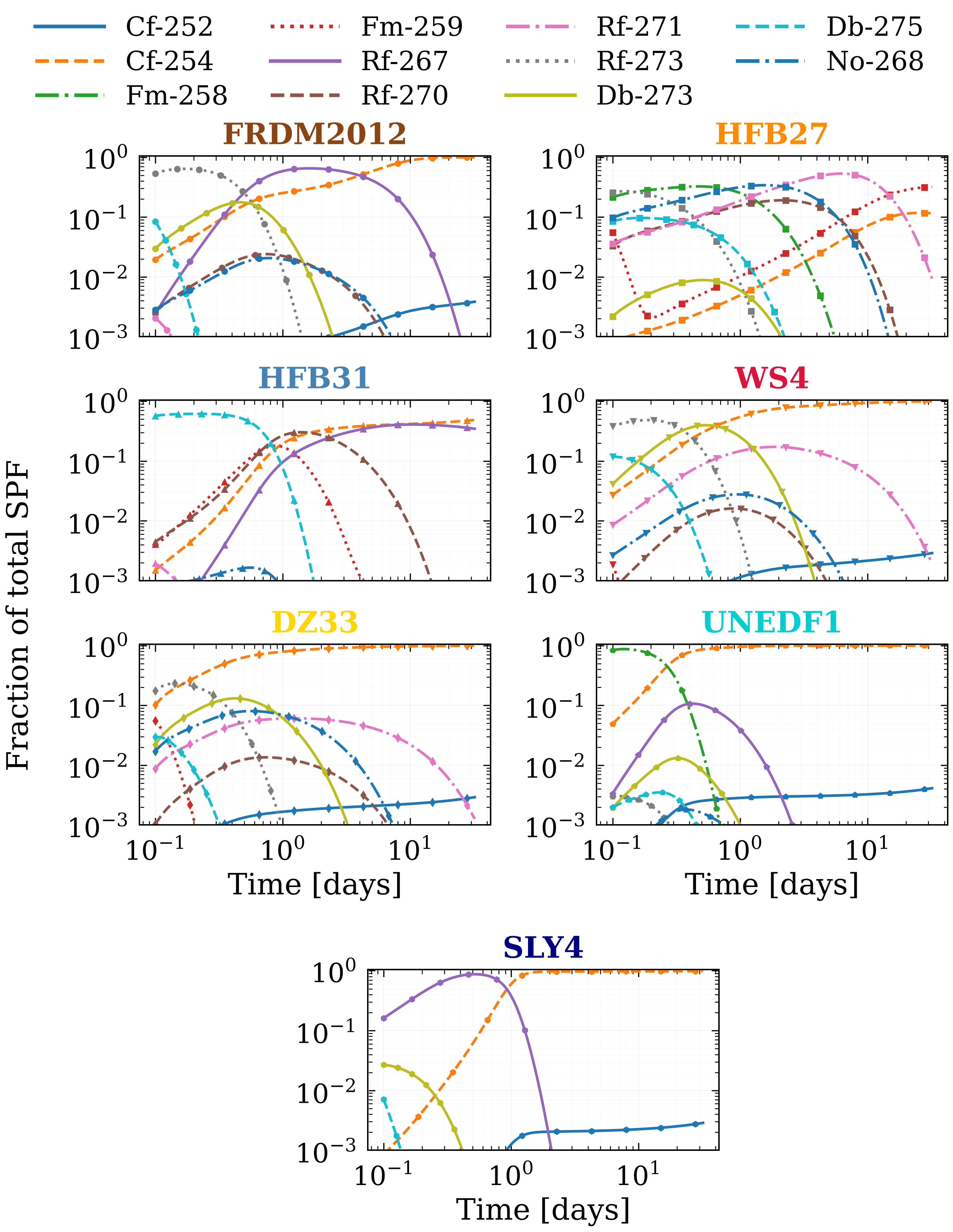}
    \caption{Per-nuclide contributions to the total spontaneous fission heating rate. \isotope[254]{Cf} dominates for most models after about 1 day, except for HFB27 which has its SPF heating driven by heavier elements such as \isotope[259]{Fm} and \isotope[268]{No}.}
    \label{spfnuclides}
\end{figure}

\begin{table*}[t]
\centering
\caption{Dominant isotope contributors to SPF heating ($>$10\% of total) at different epochs for each nuclear mass model.}
\label{tab:isotope_contributions}
\small
\setlength{\tabcolsep}{4pt}
\begin{tabular}{@{}lccccc@{}}
\toprule
\textbf{Model} & \textbf{0.5 days} & \textbf{1 day} & \textbf{5 days} & \textbf{15 days} & \textbf{30 days} \\
\midrule
FRDM2012 & \begin{tabular}[t]{@{}c@{}}\textcolor{blue!70!black}{\isotope[254]{Cf}} (26\%) \\ \isotope[267]{Rf} (23\%) \\ \isotope[273]{Rf} (22\%) \\ \isotope[273]{Db} (15\%)\end{tabular} & \begin{tabular}[t]{@{}c@{}}\isotope[267]{Rf} (49\%) \\ \textcolor{blue!70!black}{\isotope[254]{Cf}} (38\%)\end{tabular} & \begin{tabular}[t]{@{}c@{}}\textcolor{blue!70!black}{\isotope[254]{Cf}} (71\%) \\ \isotope[267]{Rf} (28\%)\end{tabular} & \begin{tabular}[t]{@{}c@{}}\textcolor{blue!70!black}{\isotope[254]{Cf}} (98\%)\end{tabular} & \begin{tabular}[t]{@{}c@{}}\textcolor{blue!70!black}{\isotope[254]{Cf}} (99\%)\end{tabular} \\[0.5ex]
\midrule
HFB27 & \begin{tabular}[t]{@{}c@{}}\isotope[258]{Fm} (47\%) \\ \isotope[268]{No} (20\%)\end{tabular} & \begin{tabular}[t]{@{}c@{}}\isotope[258]{Fm} (38\%) \\ \isotope[268]{No} (28\%) \\ \isotope[271]{Rf} (14\%) \\ \isotope[270]{Rf} (12\%)\end{tabular} & \begin{tabular}[t]{@{}c@{}}\isotope[271]{Rf} (45\%) \\ \isotope[268]{No} (14\%) \\ \isotope[270]{Rf} (11\%)\end{tabular} & \begin{tabular}[t]{@{}c@{}}\isotope[259]{Fm} (26\%) \\ \isotope[271]{Rf} (16\%) \\ \isotope[266]{No} (15\%) \\ \textcolor{blue!70!black}{\isotope[254]{Cf}} (15\%) \\ \isotope[260]{Fm} (14\%)\end{tabular} & \begin{tabular}[t]{@{}c@{}}\isotope[259]{Fm} (33\%) \\ \isotope[266]{No} (17\%) \\ \isotope[260]{Fm} (17\%) \\ \textcolor{blue!70!black}{\isotope[254]{Cf}} (16\%)\end{tabular} \\[0.5ex]
\midrule
HFB31 & \begin{tabular}[t]{@{}c@{}}\isotope[275]{Db} (38\%) \\ \isotope[259]{Fm} (20\%)\end{tabular} & \begin{tabular}[t]{@{}c@{}}\isotope[259]{Fm} (27\%) \\ \textcolor{blue!70!black}{\isotope[254]{Cf}} (26\%) \\ \isotope[270]{Rf} (22\%)\end{tabular} & \begin{tabular}[t]{@{}c@{}}\textcolor{blue!70!black}{\isotope[254]{Cf}} (54\%) \\ \isotope[267]{Rf} (28\%)\end{tabular} & \begin{tabular}[t]{@{}c@{}}\textcolor{blue!70!black}{\isotope[254]{Cf}} (59\%) \\ \isotope[267]{Rf} (29\%)\end{tabular} & \begin{tabular}[t]{@{}c@{}}\textcolor{blue!70!black}{\isotope[254]{Cf}} (63\%) \\ \isotope[267]{Rf} (25\%)\end{tabular} \\[0.5ex]
\midrule
WS4 & \begin{tabular}[t]{@{}c@{}}\textcolor{blue!70!black}{\isotope[254]{Cf}} (36\%) \\ \isotope[273]{Db} (35\%) \\ \isotope[273]{Rf} (14\%)\end{tabular} & \begin{tabular}[t]{@{}c@{}}\textcolor{blue!70!black}{\isotope[254]{Cf}} (61\%) \\ \isotope[273]{Db} (21\%) \\ \isotope[271]{Rf} (13\%)\end{tabular} & \begin{tabular}[t]{@{}c@{}}\textcolor{blue!70!black}{\isotope[254]{Cf}} (90\%)\end{tabular} & \begin{tabular}[t]{@{}c@{}}\textcolor{blue!70!black}{\isotope[254]{Cf}} (97\%)\end{tabular} & \begin{tabular}[t]{@{}c@{}}\textcolor{blue!70!black}{\isotope[254]{Cf}} (99\%)\end{tabular} \\[0.5ex]
\midrule
SLy4 & \begin{tabular}[t]{@{}c@{}}\isotope[267]{Rf} (85\%)\end{tabular} & \begin{tabular}[t]{@{}c@{}}\textcolor{blue!70!black}{\isotope[254]{Cf}} (71\%) \\ \isotope[267]{Rf} (27\%)\end{tabular} & \begin{tabular}[t]{@{}c@{}}\textcolor{blue!70!black}{\isotope[254]{Cf}} (98\%)\end{tabular} & \begin{tabular}[t]{@{}c@{}}\textcolor{blue!70!black}{\isotope[254]{Cf}} (98\%)\end{tabular} & \begin{tabular}[t]{@{}c@{}}\textcolor{blue!70!black}{\isotope[254]{Cf}} (99\%)\end{tabular} \\[0.5ex]
\midrule
UNEDF1 & \begin{tabular}[t]{@{}c@{}}\textcolor{blue!70!black}{\isotope[254]{Cf}} (83\%) \\ \isotope[267]{Rf} (12\%)\end{tabular} & \begin{tabular}[t]{@{}c@{}}\textcolor{blue!70!black}{\isotope[254]{Cf}} (94\%)\end{tabular} & \begin{tabular}[t]{@{}c@{}}\textcolor{blue!70!black}{\isotope[254]{Cf}} (100\%)\end{tabular} & \begin{tabular}[t]{@{}c@{}}\textcolor{blue!70!black}{\isotope[254]{Cf}} (100\%)\end{tabular} & \begin{tabular}[t]{@{}c@{}}\textcolor{blue!70!black}{\isotope[254]{Cf}} (100\%)\end{tabular} \\[0.5ex]
\midrule
DZ33 & \begin{tabular}[t]{@{}c@{}}\textcolor{blue!70!black}{\isotope[254]{Cf}} (80\%)\end{tabular} & \begin{tabular}[t]{@{}c@{}}\textcolor{blue!70!black}{\isotope[254]{Cf}} (90\%)\end{tabular} & \begin{tabular}[t]{@{}c@{}}\textcolor{blue!70!black}{\isotope[254]{Cf}} (98\%)\end{tabular} & \begin{tabular}[t]{@{}c@{}}\textcolor{blue!70!black}{\isotope[254]{Cf}} (99\%)\end{tabular} & \begin{tabular}[t]{@{}c@{}}\textcolor{blue!70!black}{\isotope[254]{Cf}} (100\%)\end{tabular} \\[0.5ex]
\bottomrule
\end{tabular}
\end{table*}

\subsection{Per-nuclide Variations in the Spontaneous Fission Heating Rate}\label{subsec:spfnuc}
The radiative-transfer observables are sensitive to the heating rate, which sets the overall luminosity scale, so the heating-rate differences between mass models should produce a corresponding spread in the light curves. We therefore examine the two channels in which the mass models differ most, beginning with spontaneous fission. Fig.~\ref{spfnuclides} decomposes the SPF heating rate of each model by nuclide, and Tab.~\ref{tab:isotope_contributions} lists every nuclide contributing more than 10\% of the SPF heating at selected epochs (half-lives and fission-rate sources are given in Tab.~\ref{tab:isotope_properties}). By 15~days, \isotope[254]{Cf} supplies $\geq 97\%$ of the SPF heating in every model except HFB27 and HFB31. Before then, and in those two models throughout, the decomposition differs considerably between models.

DZ33 and UNEDF1 are \isotope[254]{Cf}-dominated at every tabulated epoch ($\geq 80\%$), and SLy4 is from 1~day onward. Their early-time differences come from short-lived carriers. In SLy4, \isotope[267]{Rf} supplies 85\% of the SPF heating at 0.5~days before falling to 27\% at 1~day as \isotope[254]{Cf} takes over. In UNEDF1, \isotope[258]{Fm} dominates only before 0.5~days, after which \isotope[254]{Cf} carries the heating with a smaller \isotope[267]{Rf} contribution (12\% at 0.5~days). These carriers are visible in the total SPF curves (Fig.~\ref{fig:spfheating}). SLy4 lies well above UNEDF1 until the two cross near 1~day, consistent with the 150-minute half-life of \isotope[267]{Rf}, which releases its fission energy on an hours-long timescale. UNEDF1's early \isotope[258]{Fm} contribution leaves little imprint because of its 370~$\mu$s half-life. DZ33 lies between SLy4 and UNEDF1 until ${\sim}0.75$~days, then plateaus above both for the rest of the time range. Because \isotope[254]{Cf} supplies $\geq 90\%$ of DZ33's SPF heating from 1~day onward, and no other nuclide exceeds 10\% at any epoch, this offset must come from a larger absolute \isotope[254]{Cf} heating rate. We attribute it to greater production of the $A = 254$ isobar in DZ33 \citep{Zhu_2018}.

The remaining models reach \isotope[254]{Cf} dominance later. At 0.5~days, FRDM2012's SPF heating is shared among \isotope[254]{Cf} (26\%), \isotope[267]{Rf} (23\%), \isotope[273]{Rf} (22\%), and \isotope[273]{Db} (15\%). \isotope[267]{Rf} then peaks at 49\% at 1~day before \isotope[254]{Cf} takes over (71\% at 5~days, 98\% at 15~days). Both Rf isotopes have half-lives of ${\sim}150$--$170$ minutes and so shape the curve on hours-long timescales. WS4 is similar but with \isotope[273]{Db} in place of \isotope[267]{Rf} at early times (35\% at 0.5~days), and it is \isotope[254]{Cf}-dominated (90\%) by 5~days. HFB31's early SPF heating is led by \isotope[275]{Db} (38\% at 0.5~days; $t_{1/2} = 561$~s), which gives it a high initial rate. By 1~day the heating is shared among \isotope[259]{Fm}, \isotope[254]{Cf}, and \isotope[270]{Rf}. HFB31 then runs below and roughly parallel to FRDM2012 but never becomes fully \isotope[254]{Cf}-dominated: \isotope[254]{Cf} supplies 54-63\% from 5~days onward, and \isotope[267]{Rf} retains 25-29\% out to 30~days.

HFB27 is the outlier. It has the highest SPF heating rate at all epochs, a distinct late-time slope, and \isotope[254]{Cf} never exceeds 16\% of its SPF heating. Its carriers instead form a time-ordered sequence of heavy nuclides: \isotope[258]{Fm} and \isotope[268]{No} at $\lesssim 1$~day, \isotope[271]{Rf} (45\% at 5~days), and \isotope[259]{Fm}, \isotope[266]{No}, and \isotope[260]{Fm} from 15~days onward. This follows directly from HFB27's much larger $X_{Z \geq 100}$ (Figs.~\ref{fig: abundances} and \ref{fig: xlanxactmw}).
Several of these carriers have half-lives far shorter than the times at which they contribute. \isotope[258]{Fm} ($t_{1/2} = 370\,\mu$s) remains present to ${\sim}4$~days, and \isotope[259]{Fm} (1.5~s) and \isotope[260]{Fm} (4~ms) grow in importance at late times. \isotope[267]{Rf} in HFB31 behaves the same way. Such nuclides must be continuously fed by longer-lived precursors, so their heating is released on the precursors' timescales; we return to this in Sec.~\ref{sec:conclusions}. 
Because a nuclide's SPF power scales with its abundance, decay rate, and fission $Q$-value, even suitably long-lived fissioning nuclei matter only if they are produced in quantity. HFB27's enhanced production keeps its SPF heating elevated after shorter-lived or less abundant contributors have decayed. Since SPF thermalizes efficiently over this timescale (Fig.~\ref{thermeff}), these differences should propagate strongly into the light curves and spectra. The $\alpha$-decay heating also varies significantly between models (Fig.~\ref{fig:frac-heating}), and we examine it in the next section.

\begin{table}[t]
\centering
\caption{Nuclear properties of dominant SPF heating isotopes. SPF HL is the spontaneous fission half-life. We include \isotope[269]{Rf} despite it not showing up as a dominant fissioning isotope for any models, as the half-life we acquire for it using our prescription is significantly longer than that found in \cite{Barnes_2021}. We also denote the fission rate (FR) used for each isotope, with NB corresponding to the NuBase experimental data and KZ corresponding to the \cite{Karpov_2012} theoretical spontaneous fission rates.}
\label{tab:isotope_properties}
\small
\setlength{\tabcolsep}{6pt}
\begin{tabular}{@{}lccc@{}}
\toprule
\textbf{Isotope} & \textbf{SPF HL} & \textbf{FR} & \\
\midrule
\isotope[254]{Cf} & 60.7 d & NB\\
\isotope[273]{Db} & 492 min & KZ\\
\isotope[275]{Db} & 561 s & KZ\\
\isotope[258]{Fm} & 370 $\mu$s & NB\\   
\isotope[259]{Fm} & 1.5 s & NB\\
\isotope[260]{Fm} & 4 ms & NB\\
\isotope[266]{No} & 140 d & KZ\\
\isotope[268]{No} & 23.7 h & KZ\\
\isotope[267]{Rf} & 150 min & NB\\
\isotope[269]{Rf} & 3.07 yr & KZ\\
\isotope[270]{Rf} & 32.7 h & KZ\\
\isotope[271]{Rf} & 113 h & KZ\\
\isotope[273]{Rf} & 168 min & KZ\\
\bottomrule
\end{tabular}
\end{table}

\begin{table*}[htbp]
\centering
\caption{Mass fractions for the most active alpha decay channel elements Z=84 (Po), Z=86 (Rn), and Z=88 (Ra) ranked by decreasing total mass fraction.}
\label{tab:mass_fractions_z84_86_88}
\small
\begin{tabular}{@{}clcccc@{}}
\toprule
\textbf{Model} & \textbf{Z=84 (Po)} & \textbf{Z=86 (Rn)} & \textbf{Z=88 (Ra)} & \textbf{Total} \\
\midrule
DZ33         & $5.05\times10^{-7}$ & $1.75\times10^{-3}$ & $1.09\times10^{-2}$ & $1.26\times10^{-2}$ \\
HFB27        & $2.55\times10^{-7}$ & $8.75\times10^{-4}$ & $7.29\times10^{-3}$ & $8.17\times10^{-3}$ \\
FRDM2012     & $2.11\times10^{-7}$ & $7.17\times10^{-4}$ & $5.07\times10^{-3}$ & $5.79\times10^{-3}$ \\
UNEDF1       & $2.06\times10^{-7}$ & $7.05\times10^{-4}$ & $4.16\times10^{-3}$ & $4.86\times10^{-3}$ \\
WS4          & $1.76\times10^{-7}$ & $5.95\times10^{-4}$ & $4.18\times10^{-3}$ & $4.77\times10^{-3}$ \\
HFB31        & $1.93\times10^{-7}$ & $6.83\times10^{-4}$ & $3.86\times10^{-3}$ & $4.54\times10^{-3}$ \\
SLy4         & $2.72\times10^{-8}$ & $9.24\times10^{-5}$ & $5.93\times10^{-4}$ & $6.86\times10^{-4}$ \\
\bottomrule
\end{tabular}
\end{table*}

\subsection{Per-nuclide Variations in Alpha Decay Heating Rate}\label{subsec:alfnuc}
In this section we analyze the differences in the alpha decay heating rates between mass models in greater detail. Fig. \ref{fig:frac-heating} shows that DZ33 and HFB27 have a significantly higher portion of their total heating dedicated to alpha decay and SLy4 has a significantly lower portion. The other models cluster together and show little significant deviation. We attribute this to the differing mass fractions of \isotope[212]{Po}, \isotope[220]{Rn}, and \isotope[224]{Ra}, produced by each model which are the three isotopes that we find contribute most to the alpha decay heating rate across all models. Tab. \ref{tab:mass_fractions_z84_86_88} ranks each model by its total mass fraction of these three elements, with DZ33 exhibiting a much higher mass fraction and SLy4 exhibiting a much lower one. All three of these isotopes have alpha branching ratios of essentially 100 percent from NuBase2020 \citep{Kondev_2021, IAEA_NUBASE} and thus are primarily responsible for the differences seen in the alpha decay heating rates between mass models. Based on these results, we expect the alpha decay heating rate to have a notable impact on the observables of DZ33, HFB27, and SLy4 in particular. But, due to alpha decay heating's lower thermalization efficiency, it will have a much smaller impact on the observable differences across models than SPF heating.

\begin{figure}
    \centering
    \includegraphics[angle=0,width =1\linewidth]{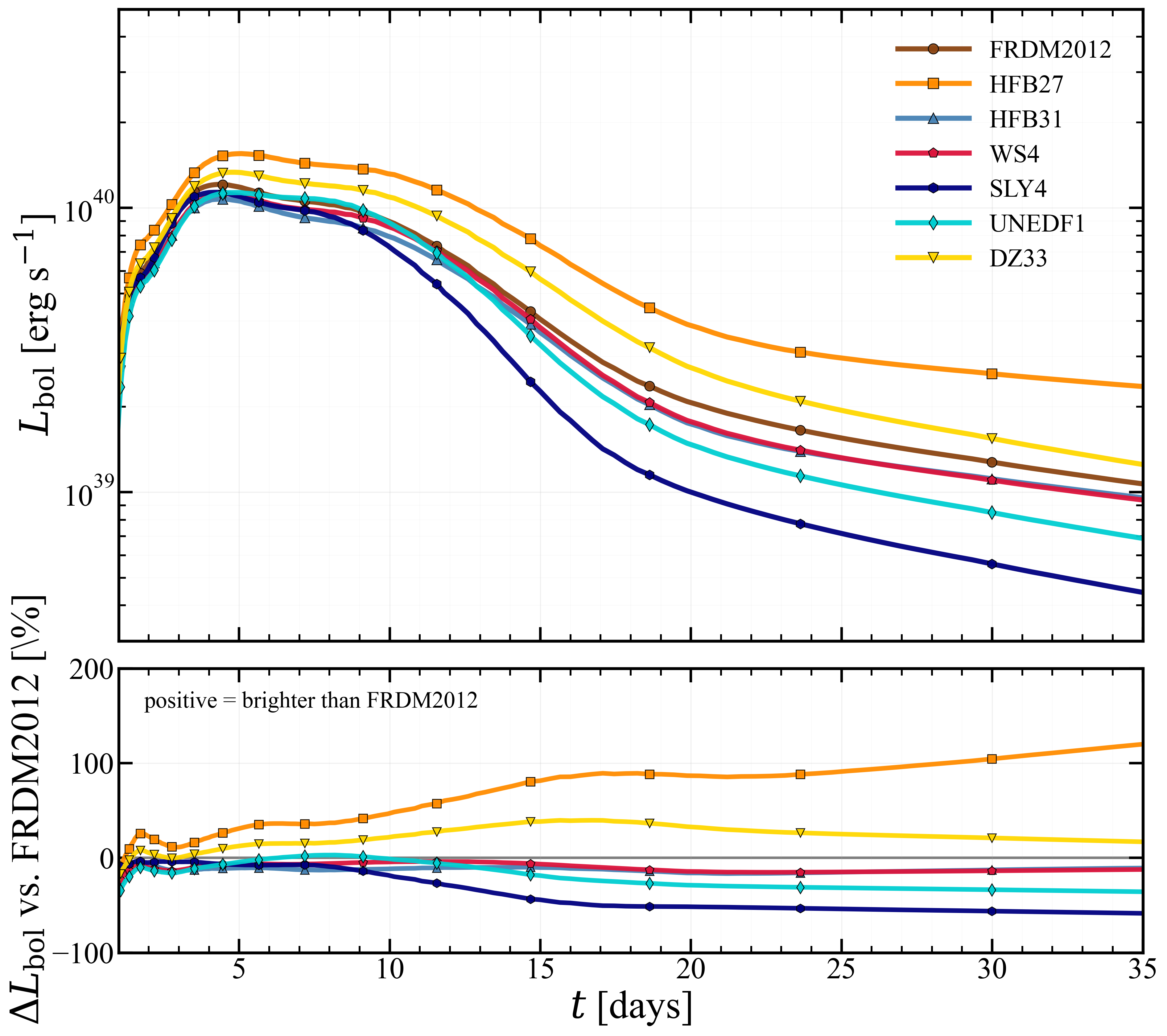}
    \caption{Bolometric luminosity curves for all nuclear mass models. The bottom panel shows the bolometric luminosity of each model relative to the FRDM2012 model. HFB27 is uniformly brighter than every other model owing to its excess of $Z\geq100$ elements driving its increased spontaneous fission heating rate. Overall, there is a considerable spread in the bolometric luminosities across this mass-weighted $Y_e$ ensemble, but much less so than for a single $Y_e$ trajectory.}
    \label{nmmbolos}
\end{figure}

\subsection{Nuclear Mass Model Variations in Light Curves}\label{subsec:nmmvlc}
\begin{figure}[h!]
    \centering
    \includegraphics[angle=0,width =1\linewidth]{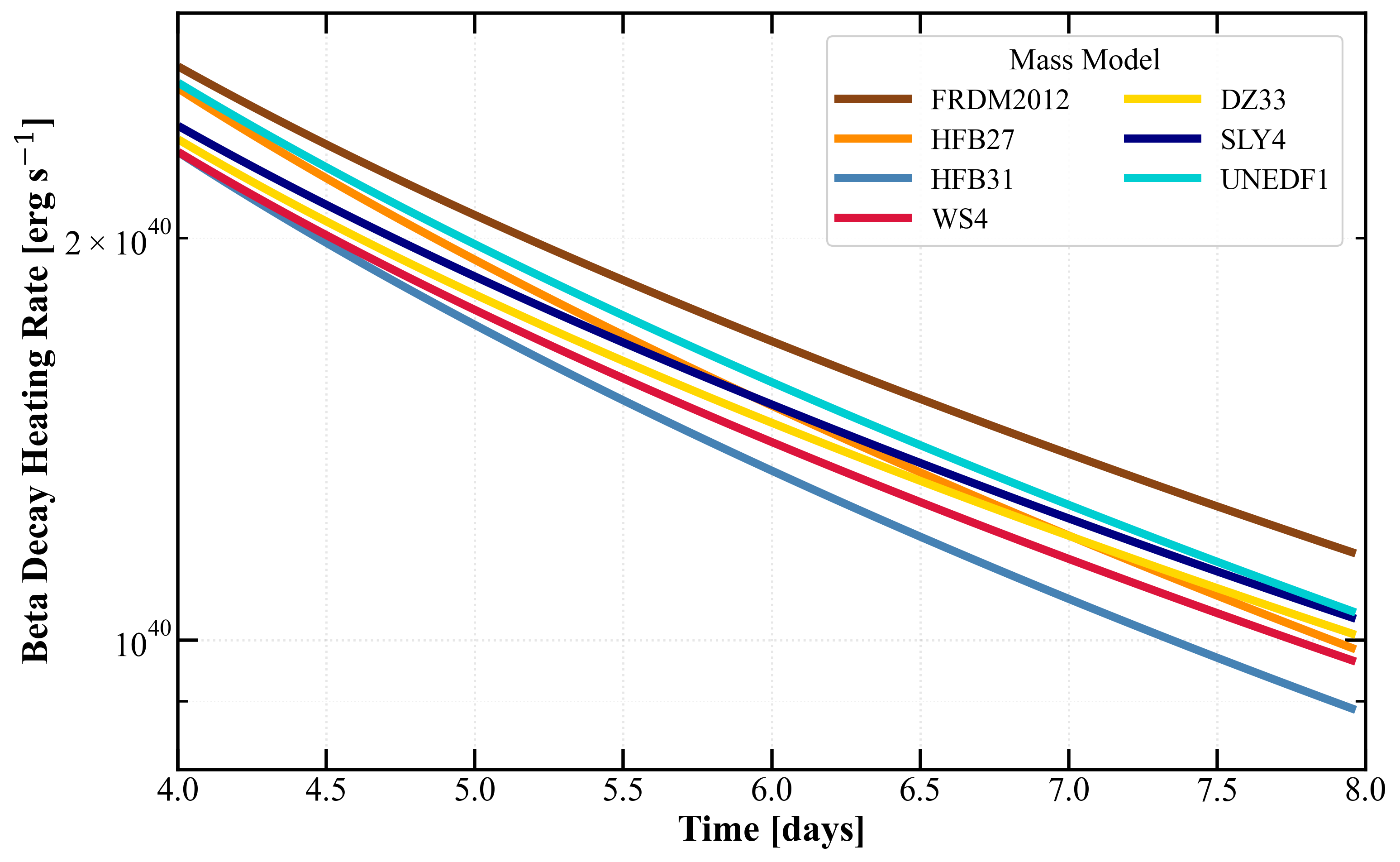}
    \caption{Beta decay heating rates over the time range where all of the mass models peak in bolometric luminosity. The rates begin to diverge starting in this epoch, which drives the dimmer luminosity peaking of HFB31.}
    \label{betadecayheatingpeakbolo}
\end{figure}

\begin{figure}[h!]
    \centering
    \includegraphics[angle=0,width =1\linewidth]{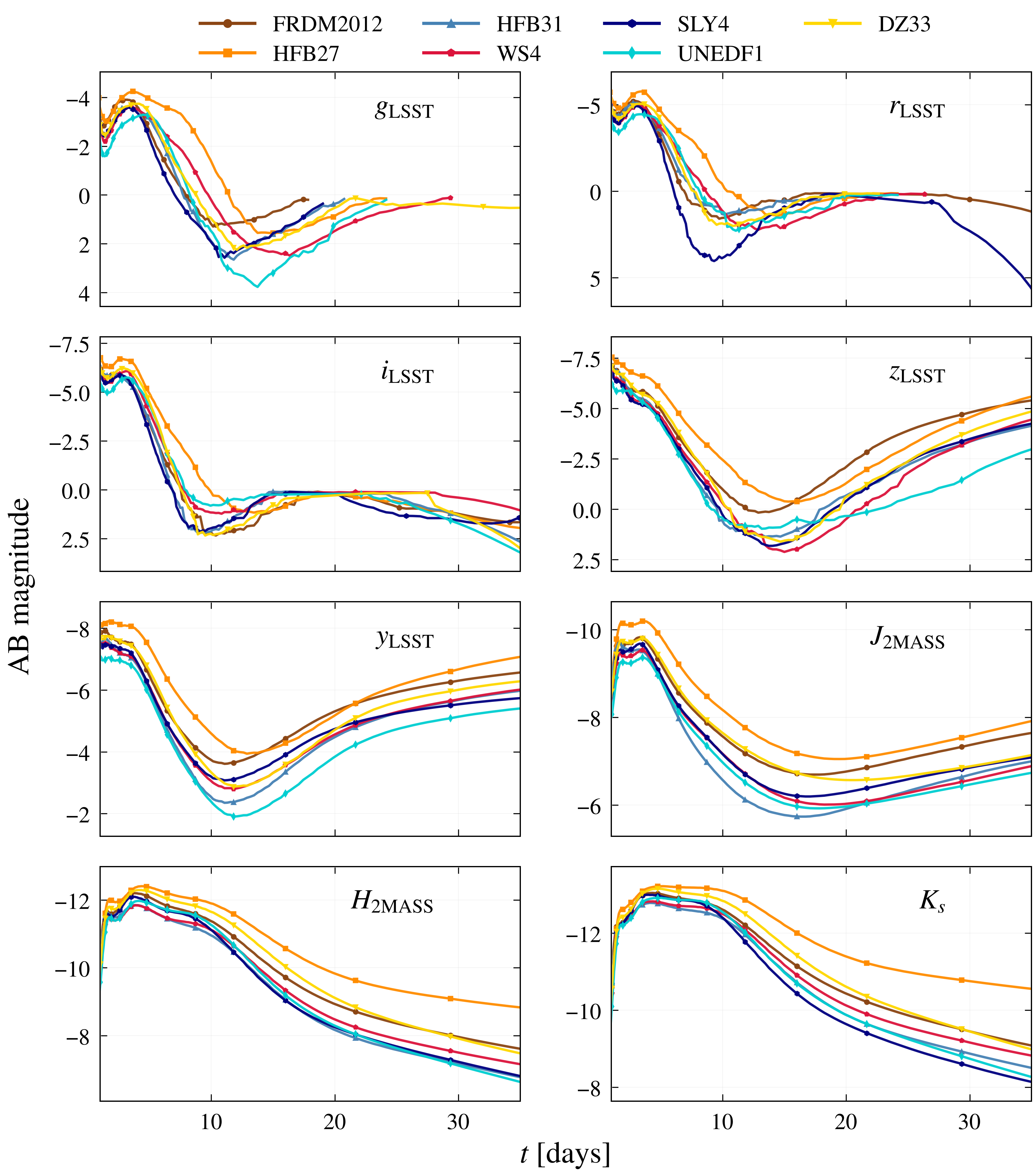}
    \caption{Multi-band light curves across eight distinct bandpasses. We use $grizy$ filters from the Legacy Survey of Space and Time (LSST), J and H filters from 2MASS to capture infrared behavior, and a Ks band from MOSFIRE.}
    \label{nmmmulti}
\end{figure}

We now examine how nuclear mass model variations propagate into bolometric and multi-band light curves. As discussed in Subsecs. \ref{subsec:spfnuc} and \ref{subsec:alfnuc} we expect the bulk of the difference in the brightness to be driven by differences in spontaneous fission heating due to its higher thermalization efficiency, modulated by outliers in alpha decay heating. We see in Fig. \ref{nmmbolos} that even when the total effective heating rates of each model are similar across tracers, there is still an approximately $45\%$ difference in peak bolometric luminosity across mass models. The ordering of the brightness of the light curves past peak luminosity roughly follows the ordering of models in their SPF heating rate, except for DZ33. This is likely due to DZ33's much higher alpha decay heating rate relative to the other models as shown in Subsec. \ref{subsec:alfnuc}. Based off our analysis, we know the specific isotopes causing much of the difference in brightness between curves as given in Tabs. \ref{tab:isotope_contributions} and \ref{tab:isotope_properties}. In the case of HFB27, its elevated alpha decay and SPF decay heating rates are causing it to be a large outlier in the spread, while SLy4 has the opposite problem. We also note that the time that each light curve peaks does not vary much across mass models in our setup. Our peak bolometric luminosities are all on the order of $10^{40}\:\text{erg s}^{-1}$, lower than typical literature values on the order of $10^{41}\:\text{erg s}^{-1}$. This is due to our ejecta mass ($6.84\times10^{-3}\: M_{\odot}$) being an order of magnitude smaller than values around $5.0\times10^{-2}\: M_{\odot}$ used in studies like \cite{Zhu_2021} and \cite{Barnes_2021}, and the fact that peak luminosity is roughly proportional to ejecta mass. We see interesting behavior in the peak luminosity values of each model relative to FRDM2012 (Appendix Tab. \ref{tab:lcstats}). HFB31 has the dimmest peak bolometric luminosity. This cannot be explained simply by alpha or SPF decay heating rates, as those of SLy4 and UNEDF1 are much lower in this time range. If we look at the beta decay heating rates at the time range where the light curves peak (Fig. \ref{betadecayheatingpeakbolo}), we see that HFB31 has the lowest beta decay heating rate across this time scale. HFB31's SPF decay heating rate is approximately constant across this time, and its alpha decay heating rate is increasing. Thus, HFB31 peaks dimmer due to this lower beta decay rate around the diffusion time, not because of alpha or SPF decay heating. FRDM2012's relatively early peak is consistent with its lower opacity owing to a lack of elements with $Z\geq100$ partially canceling with its high beta decay heating rate at this time. Turning to the multi-band light curves in Fig. \ref{nmmmulti}, the most notable feature is a rise in the AB magnitude past about 10 days in the near-infrared bands of $z_{\rm LSST}$ through $J_{\rm 2MASS}$ (filter wavelength ranges given in \citealt{Ivezic_2019, Skrutskie_2006, McLean_2012}). This rise is strongest in the $z_{\rm LSST}$ band and tapers off as the bands get redder. This supports the well-known result that kilonovae release much of their radiation in the near-infrared, and that lanthanide and actinide rich ejecta produce redder light curves and spectra \citep{Tanaka_2013, Metzger_2014, Tanaka_2020}. Optical radiation drops off rapidly and stays low, while the reddest bands ($H_{\rm 2MASS}$ and $K_s$) have the most negative magnitude and monotonically decrease across the timescale of interest. Looking at differences between models, there is little distinct ordering in the bluer bands, but the distinction between models grows as the throughput goes redward. In the reddest bands, the ordering of the magnitudes closely follows the ordering of the bolometric light curves, with HFB27 being the brightest across time and SLy4 being the least. The ejecta is optically thick in the blue bands due to multiple overlapping effects, including line blanketing from lanthanides and actinides and the strong dependence of opacity on temperature. On top of this, lanthanide and actinide line forests are densest at blue wavelengths \citep{Kasen_2013, Fontes_2020}, and the opacity saturates with increasing lanthanide fraction \citep{Barnes_2021}. All seven models are lanthanide-rich, so they all lie in this saturated regime. Blue opacity is therefore nearly insensitive to the differences between models, and no model stays brightest across the $g$--$i$ bands. Toward redder wavelengths the line density drops, the ejecta becomes optically thinner, and line blanketing no longer confounds the magnitudes. Photons absorbed in the blue line forest are reprocessed to longer wavelengths, so from $z$ redward the bands act as a rough bolometer.

Overall, we find a considerable, but much smaller spread in the light curves than found by \cite{Zhu_2021} and \cite{Barnes_2021} across this 30-tracer $Y_e$ composition ensemble. The ratio of the brightest to faintest bolometric peak is 1.45 for our seven mass models (Table~\ref{tab:lcstats}), about 6 in \citet{Zhu_2021}, and greater than 10 in \citet{Barnes_2021}. The earlier model suites also varied initial $Y_e$ and fission prescriptions. We explore this contrast in the discussion section.

\begin{figure*}[t]
    \centering
    \includegraphics[width=0.95\textwidth]{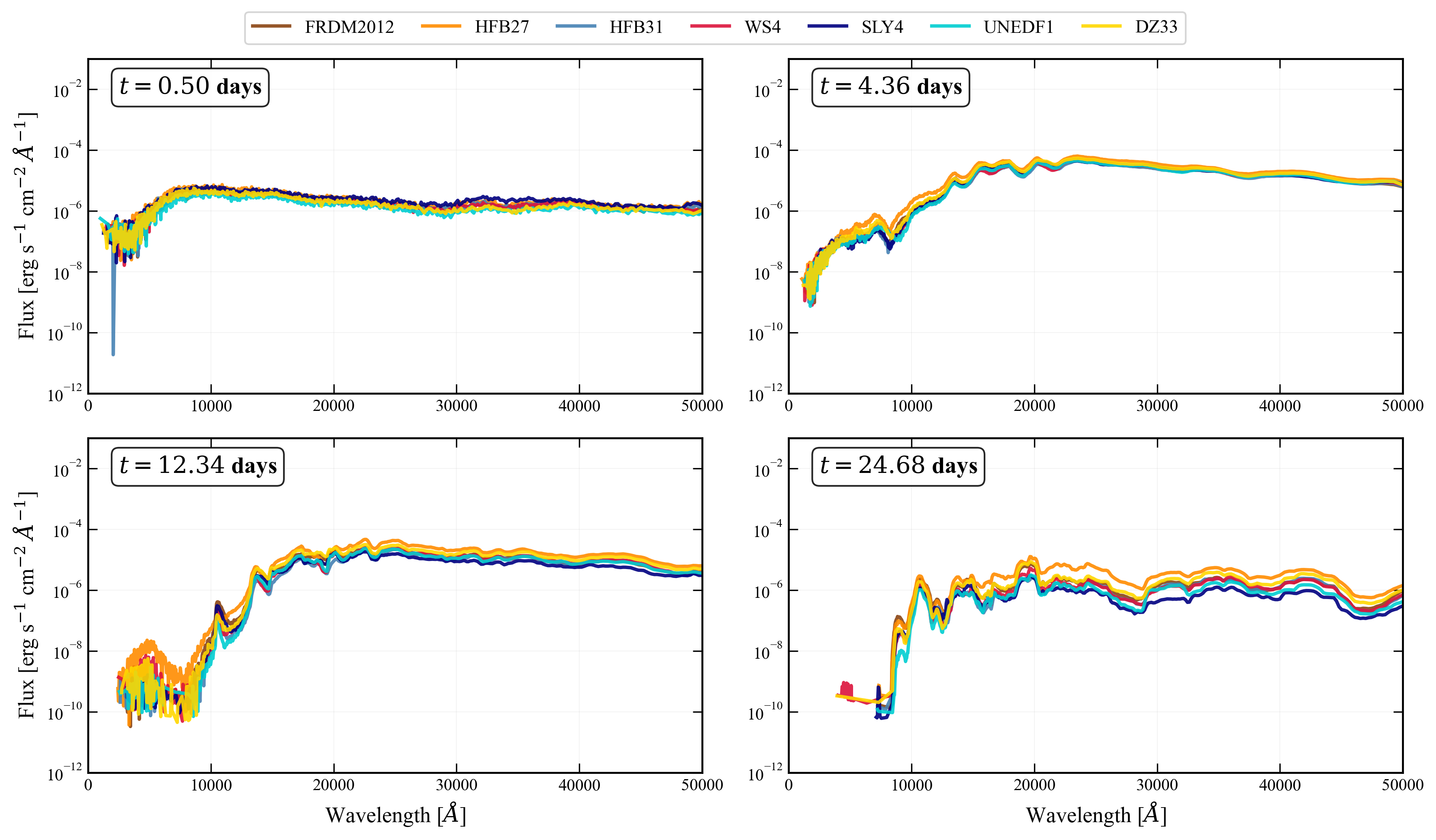}
    \caption{Spectral evolution across different time epochs for all nuclear mass models. Each panel shows spectra from all models at a specific time, demonstrating the wavelength-dependent differences between mass models and their evolution with time.}
    \label{nmmspectraall}
\end{figure*}

\subsection{Nuclear Mass Model Variations in Spectra}\label{subsec:nmmvspec}
We now turn to disentangling the effect of differing nuclear mass models on trends in the spectra also produced by \texttt{SuperNu}. There have been studies on the spectral signatures of either lanthanide and actinide rich ($X_{\rm lan,\,act}\gtrsim10^{-2}$) or lanthanide and actinide poor ($X_{\rm lan,\,act}\lesssim10^{-3}$) ejecta \citep{Metzger_2014, Tanaka_2020}, as well as on much higher $X_{\rm lan,\,act}$ ejecta \citep{Barnes_2021}. Here, we look to identify spectral trends across an ejecta cloud composed of both high and low $X_{\rm lan,\,act}$ shells, for each mass model. Based on the nuclear mass model producing the most lanthanides alternating across elements as shown in our mass-weighted abundance (Fig. \ref{fig: abundances}) and the homogeneity of the mass-weighted lanthanide mass fractions across models, as well as the similar gray opacity for each model, we expect that much of the spectral difference will be driven by, but not fully explained by, the differences seen in the actinides and superheavies where the models diverge significantly under mass-weighting (Fig. \ref{fig: xlanxactmw}). As discussed in the last section, HFB27 produces a brighter and later-peaking bolometric light curve attributed to its excess of spontaneous fission heating and slightly higher gray opacity from its heavy isotope production. Looking at Fig. \ref{nmmspectraall} we see that this trend propagates into the spectra with HFB27 boasting the brightest spectrum at all temporal epochs. We see the least deviation at early times, before the higher SPF heating rate of HFB27 causes a prolonged brightening of its observables. At later times and redder wavelengths, the flux noticeably separates between mass models and roughly follows the ordering of the bolometric light curves. It is a well-known result that kilonovae release much of their radiation in the near infra-red, and emission is generally suppressed in the optical band for earlier times \citep{Barnes_2013, Barnes_2021, Tanaka_2013, Villar_2017, Arcavi_2017}. This is consistent with the results of our spectra, as the flux peaks at most epochs (except very early ones) around $20000 \: \si{\angstrom}$. Additionally, the flux increases by nearly two orders of magnitude in the near-infrared bands while the flux decreases in the bluest bands between 0.50 days and 4.36 days as shown in Fig. \ref{nmmspectraall}.

There has been work on which elements most affect trends in kilonova spectra \citep{Kasen_2017, Even_2020}, which suggest that elements such as Nd and U have a large effect due to their complex open-shell electron structures producing dense forests of bound--bound transitions and therefore high opacities. Lanthanides in general are expected to also have an effect on the spectra by suppressing optical emission and shifting emission to redder wavelengths because their partially filled 4f electron shells produce a high density of bound--bound transitions and hence large expansion opacities \citep{Tanaka_2013, Kasen_2017}. In our mass-weighted ensemble where the lanthanide mass fraction hardly differs across mass models, we cannot attribute spectral differences to differences in $X_{\rm lan}$. Thus, the primary factor driving flux differences in the spectra likely comes from the differences in actinide and superheavy element mass fractions across models and the associated differences in the SPF heating rate creating brighter and longer-lasting radiation. As mentioned in Subsec. \ref{subsec:spfnuc}, for HFB27 in particular the isotopes dominating its SPF heating rate at later times include \isotope[259]{Fm}, \isotope[266]{No}, and \isotope[260]{Fm}, while the rest of the models are dominated almost completely by \isotope[254]{Cf}. Therefore, the separation between mass models in the late-time spectra may represent a potentially disentangleable observational signature of differing actinide and superheavy production. However, this signature should be interpreted as a signature of the heating pattern produced by groups of nuclei, rather than as a direct spectroscopic identification of individual isotopes.  In particular, models dominated by \isotope[254]{Cf} heating may exhibit different late-time decline rates and near-infrared flux evolution than models such as HFB27, where \isotope[259]{Fm}, \isotope[266]{No}, and \isotope[260]{Fm} contribute more strongly to the SPF heating. Because the mass-weighted lanthanide fractions and gray opacities are similar across the mass models, such differences are less likely to arise from lanthanide-opacity effects alone and may instead trace differences in actinide and superheavy nucleosynthesis in this mass-weighted, multi-shell varied $Y_e$ outflow. One caveat applies specifically to the late-time separation described above. The \citet{Fontes_2020, Fontes_2023} opacities assume local thermodynamic equilibrium, so ionization states and level populations follow Saha-Boltzmann at the local gas temperature. As the ejecta expands, the recombination timescale grows relative to the expansion timescale and the ionization state freezes out above its LTE value; at the same time, a growing fraction of the radioactive energy is deposited by non-thermal electrons, $\alpha$ particles, and fission fragments that ionize and excite directly rather than heating the thermal electron pool \citep{Barnes_2016, Brethauer_2025}. Our ejecta mass of $6.84\times10^{-3}\,M_{\odot}$ is roughly an order of magnitude below the values adopted in \citet{Zhu_2021} and \citet{Barnes_2021}, so this transition occurs correspondingly earlier here.

\section{Results: Electron Fraction Variations}\label{sec:compresults}

We now examine a model grid that isolates the effects of composition and radioactive heating at fixed ejecta structure. As before, all of these runs use a common ejecta mass of $M_{\rm ej}=6.84\times10^{-3}\,M_{\odot}$. The mass in each velocity branch, the density profile, and velocity grid is preserved in each run. The full grid has 15 target electron fractions spanning $0.10\leq Y_e\leq0.45$ in intervals of 0.025. Within each velocity bin, the GRMHD tracers are reweighted to achieve the desired electron fraction for that run.

Subsections \ref{subsec:yenuc}, \ref{subsec:yebolem}, \ref{subsec:yemagevl} and \ref{subsec:yespec} detail the effect of our $Y_e$ variations on nucleosynthesis, luminosity, broadband magnitudes and spectra, respectively. In Subsec.~\ref{subsec:yecompare}, we provide a brief comparison to other works.

\subsection{Nucleosynthesis Effects}\label{subsec:yenuc}

\begin{figure}
    \centering
    \includegraphics[angle=0,width =1\linewidth]{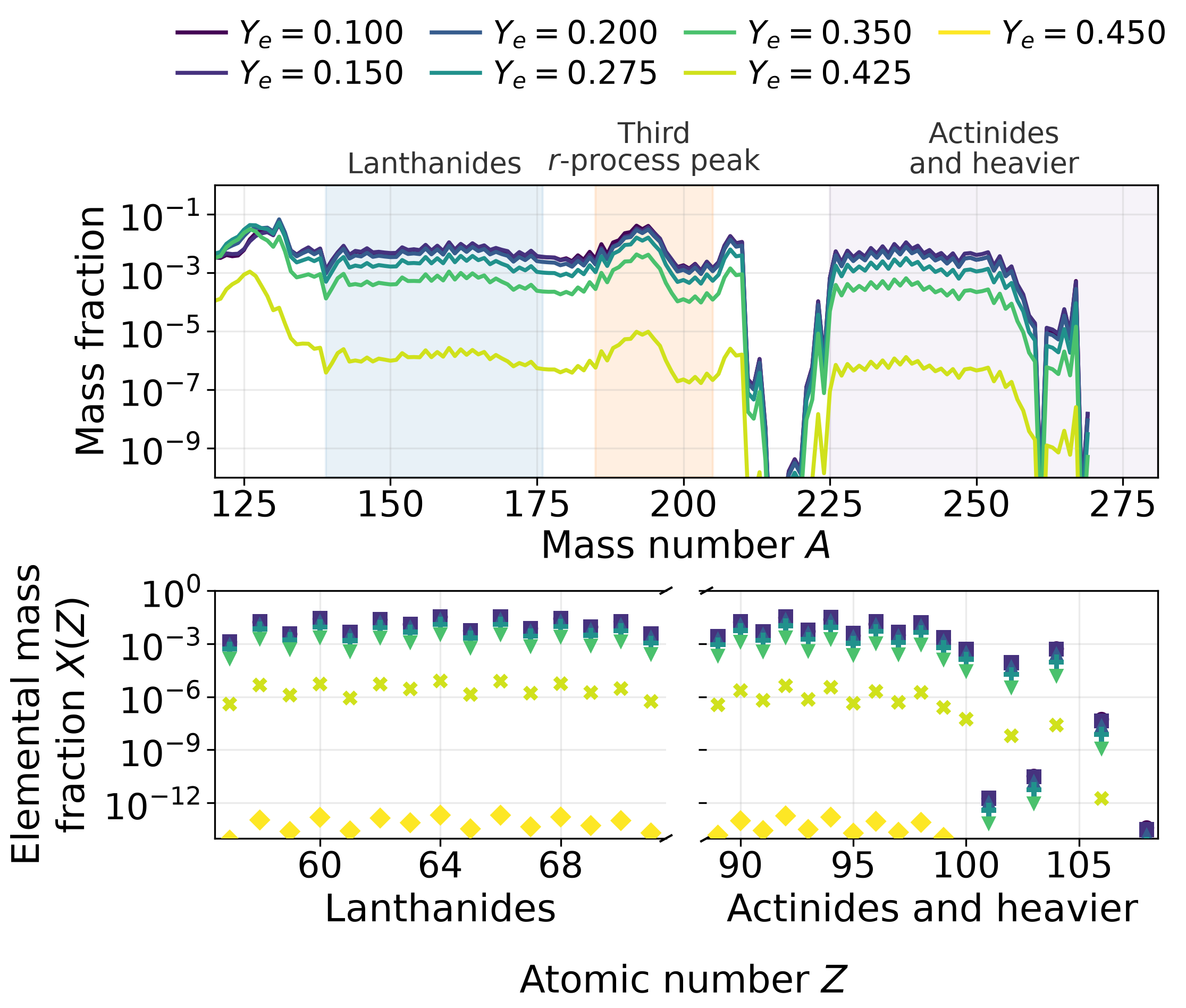}
    \caption{Mass-weighted nucleosynthetic abundance distributions at 50 days for different $Y_e$ models. The upper panel shows the heavy element ejecta abundance as a function of mass number. The lower panel shows the corresponding heavy elemental mass fraction as a function of atomic number.}
    \label{fig:ye-abundances}
\end{figure}

Fig. \ref{fig:ye-abundances} demonstrates the systematic change in the nucleosynthetic outcome as a result of a change in the ejecta electron fraction. The lower $Y_e$ models are sufficiently neutron rich and populate the third $r$-process peak and produce substantial lanthanide, actinide, and heavy-element abundances. Increasing $Y_e$ reduces the neutron-to-seed ratio and suppresses heavier nuclei. The lanthanide and actinide mass fractions decline monotonically across the grid, but the decline above $Y_e \simeq 0.325$ is particularly rapid. 

This compositional transition is important for the radiative transport calculations as heavy-element abundances affect the radioactive energy supply and the opacity of the ejecta. Particularly, the dense line structure associated with lanthanides and actinides increases the opacity and redistributes radiation towards longer wavelengths \citep{Kasen_2013, Even_2020}.

\begin{figure}
        \centering
        \includegraphics[width=1\linewidth]{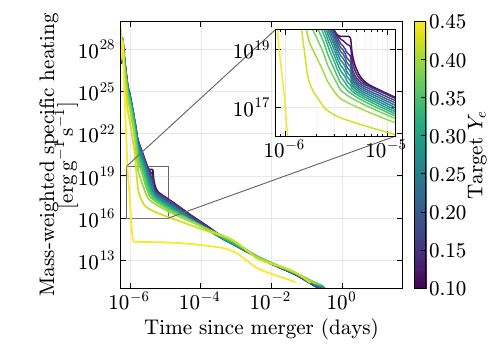}
        \caption{Mass weighted specific heating for all $Y_e$ runs. Colors denote the initial electron fraction. }
        \label{heating}
\end{figure}

\begin{figure*}[!t]
    \centering
    \begin{subfigure}{\textwidth}
        \centering
        \includegraphics[width=\linewidth]{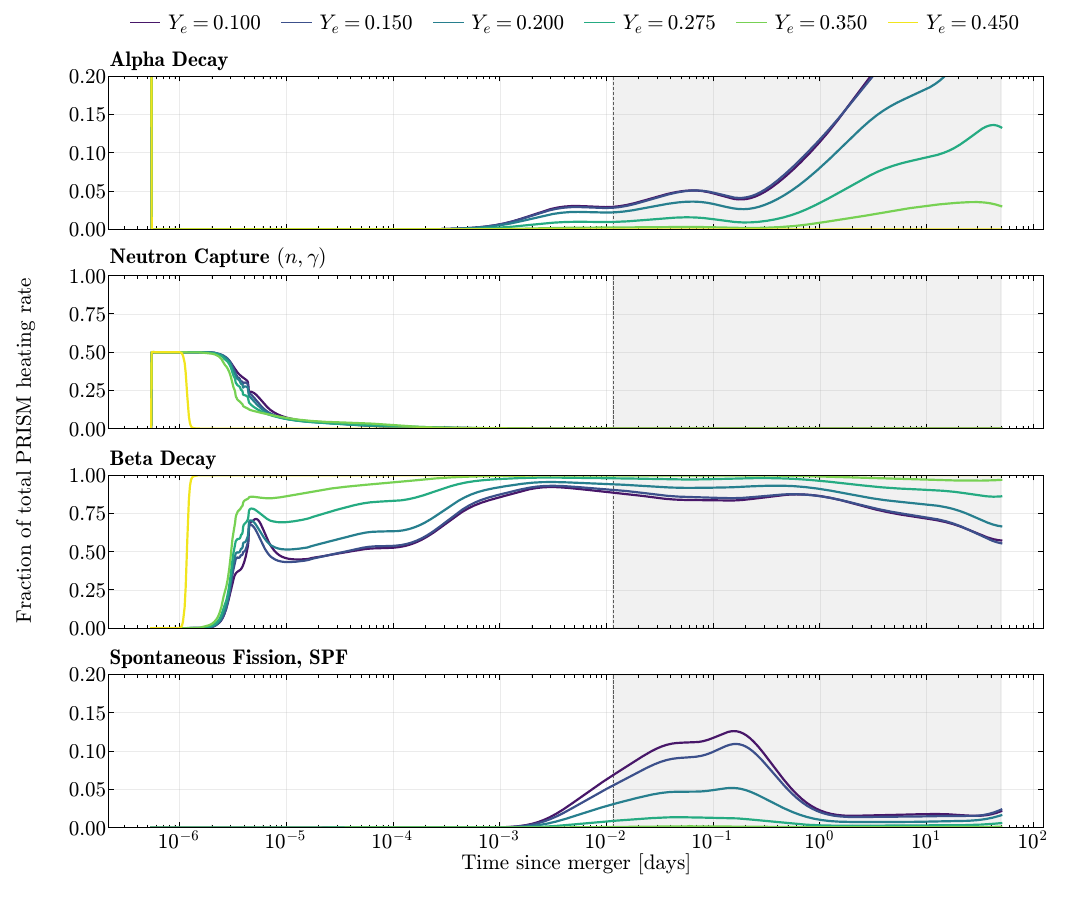}
        \caption{Fractional contribution by each process to the sum of the absolute magnitudes of the seven PRISM channel rates for representative $Y_e$. Neutron capture dominates the very earliest phase. After this phase of neutron capture freeze out, $\beta$-decay supplies a majority of the heating. As the ejecta expands and cools, the synthesized nuclei decay towards stability. The $\alpha$-decay contribution remains subdominant over the interval plotted. Its relative importance, however, increases at later times. The dashed line represents the end of the extrapolated tracer thermodynamic trajectory and the beginning of the \texttt{SuperNu} runs.}
        \label{fig:sub1}
    \end{subfigure}
    \caption{PRISM nuclear-heating decomposition for the composition variation suite. Panel (a) displays the relative channel contributions for six representative compositions. Panel (b) shows the absolute rates for the full sequence. These are PRISM nuclear energy generation rates before channel-dependent thermalization and radiative transport treatments done in \texttt{SuperNu}.}
    \label{fig:comp-heating-rates}
\end{figure*}

\begin{figure*}[!t]
    \ContinuedFloat
    \centering
    \begin{subfigure}{\textwidth}
        \centering
        \includegraphics[width=\linewidth]{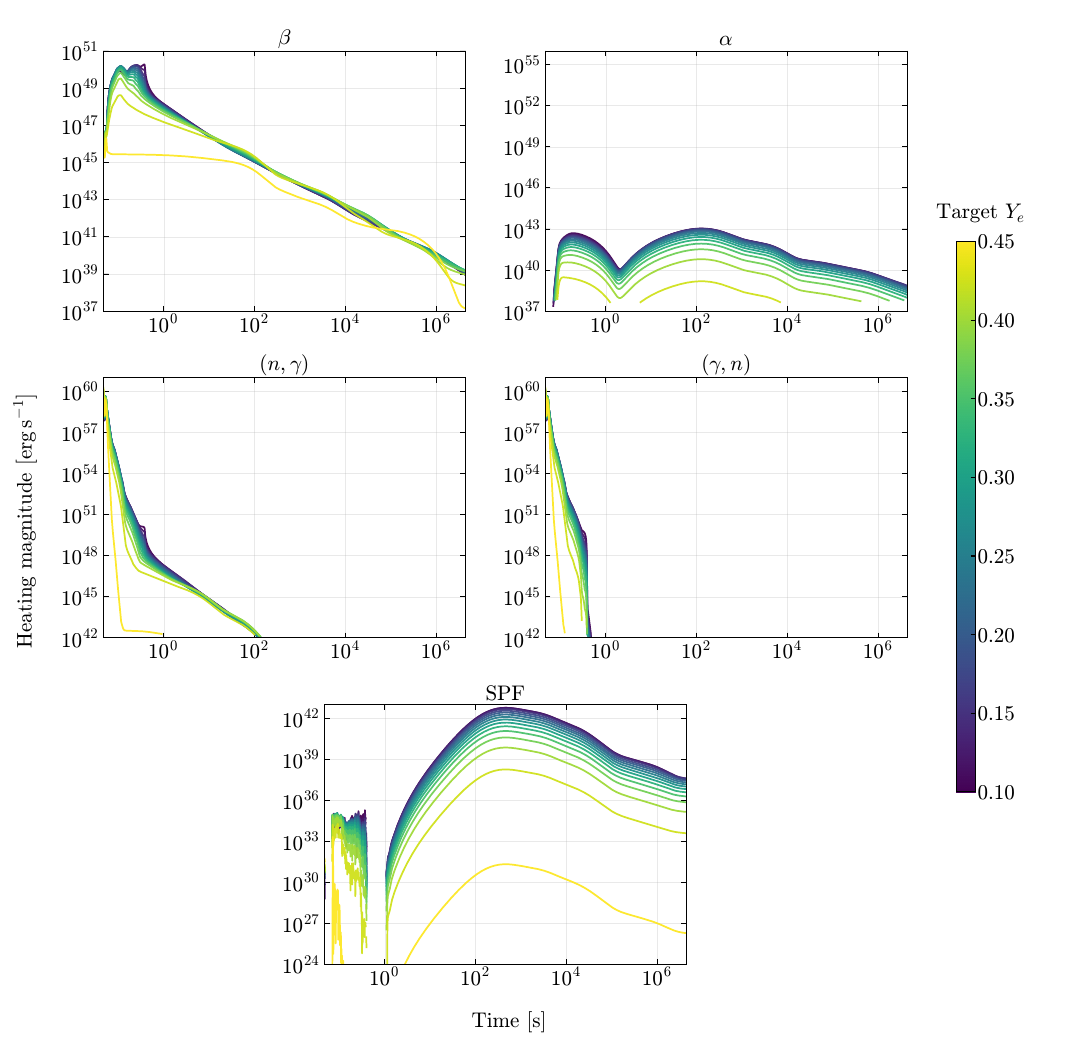}
        \caption{Ejecta-integrated absolute magnitudes of the PRISM channel rates for the full composition suite. The panels show $\beta$ decay, $\alpha$ decay, neutron capture $(n,\gamma)$, photodisintegration $(\gamma,n)$, and spontaneous fission (SPF). The $\alpha$-decay and fission rates are enhanced in the neutron-rich models since more heavy and translead nuclei are produced. }
        \label{fig:sub2}
    \end{subfigure}
    \caption[]{PRISM nuclear-heating decomposition for the composition variation suite (continued).}
\end{figure*}

\begin{figure}[h]
    \centering
    \includegraphics[width=1\linewidth]{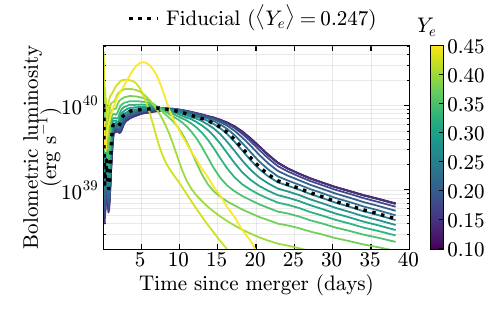}
    \caption{Bolometric luminosity as a function of time for different electron fractions. We hold fixed the ejecta mass, velocity grid, and density profile across all the runs.}
    \label{bolometric_lum}
\end{figure}

The mass weighted heating histories are shown in Fig. \ref{heating}. The models exhibit significant differences at early times when the reaction channels and short-lived nuclei produce structure in the heating curves. Most of the models between $Y_e \simeq 0.10$ and $Y_e \simeq 0.425$ subsequently approach similar declining heating trends. However, differences remain in the normalization and timing of the early heating features. The $Y_e=0.45$ model is an outlier that has a much lower heating rate over its evolution. Its behavior follows from the strong concentration of ejecta mass into the highest $Y_e$ tracers.

The channel decomposition in Fig. \ref{fig:comp-heating-rates} shows that neutron capture and photodisintegration dominate the early nuclear energy exchange rates in absolute magnitude. After this, the beta channel rapidly becomes dominant. Alpha decay remains subdominant over the timescale, but its relative contribution grows over the time examined. This late time increase is a result of the heavy progenitors that feed alpha decay chains as well as the disappearance of beta emitters \citep{Barnes_2021}.

\subsection{Bolometric Emission}\label{subsec:yebolem}

The $Y_e$ values produce large differences in both the luminosity and characteristic timescale of the kilonova. Fig. \ref{bolometric_lum} shows these variations of the bolometric luminosity for different initial electron fractions. 

\begin{figure}
    \centering
    \includegraphics[width=1\linewidth]{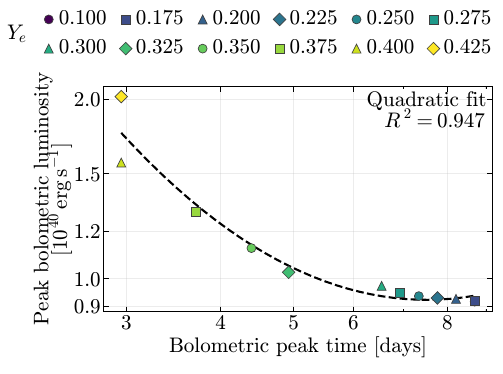}
    \caption{Peak bolometric luminosity as a function of the time of maximum luminosity for different electron fraction models. Colors denote the electron fraction value. Both quantities are measured directly from the unsmoothed \texttt{SuperNu} bolometric light curves, selecting the maximum luminosity and its corresponding time between 2 and 35 days. }
    \label{lum-vs-peak-time}
\end{figure}

The $Y_e$ values also produce differences in the characteristic timescale of the kilonova. This is demonstrated in Figs. \ref{lum-vs-peak-time} and \ref{bolometric_lum}. Over the intermediate to upper range of the initial electron fraction ($0.20 \lesssim Y_e \lesssim 0.425$), the peak luminosity increases monotonically. The measured peak rises from approximately $9.25\times10^{39}\ {\rm erg\,s^{-1}}$ for $Y_e=0.20$ to approximately $2.02\times10^{40}\ {\rm erg\,s^{-1}}$ for $Y_e=0.425$. Within this interval, the location of the light curve maximum shifts from approximately 8.2 to 3.0 days, respectively. 

The peak quantities do not fully describe the composition dependence of the emission, though. The higher electron fraction models generally produce larger early luminosities and become dimmer in later times when compared to the lower electron fractions. The substantial light curve differences among the models with similar heating demonstrate that radiative transport dynamics are not controlled solely by heating. Composition dependent opacity and diffusion determine when the deposited energy actually escapes and how it is redistributed in wavelength. The lanthanide-rich, low $Y_e$ ejecta hence produces broader and redder emission. 

\subsection{Broadband Evolution}\label{subsec:yemagevl}

\begin{figure*}
    \centering
    \includegraphics[width=1\linewidth]{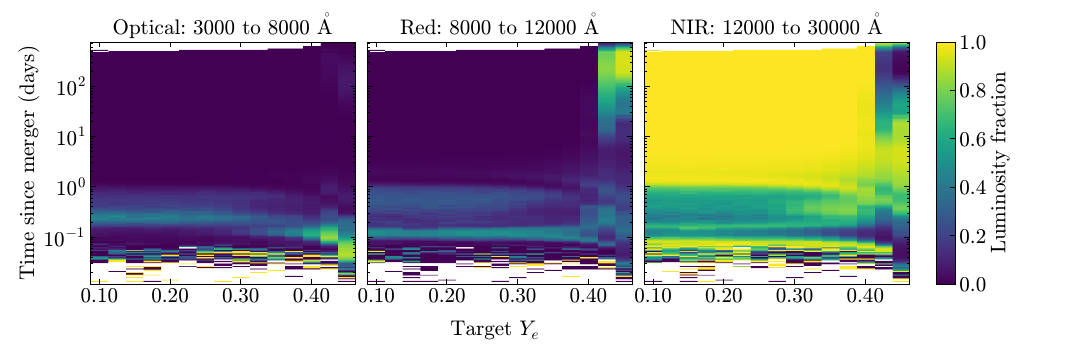}
    \caption{Fraction of the $3000$ - $30000$ \AA \space luminosity emerging in the optical ($3000$--$8000$~\AA), red ($8000$--$12000$~\AA), and near-infrared ($12000$--$30000$~\AA) intervals. The three fractions sum to unity at each time and initial $Y_e$. As shown, lower $Y_e$ runs typically enter the near-infrared dominated epoch earlier than high $Y_e$ runs.}
    \label{flux-fractions}
\end{figure*}

The broadband evolution provides a direct measure of the composition-dependent spectral redistribution. Fig. \ref{flux-fractions} shows that a substantial fraction of the early luminosity emerges in the optical band. As the ejecta expands and cools, the optical fraction generally decreases and the red and near-infrared contributions increase. The transition occurs earlier in lower-$Y_e$ models. This is likely due to the larger fraction of lanthanides and actinides that redistribute radiation towards longer wavelengths more efficiently. 

\begin{figure}
    \centering
    \includegraphics[width=1\linewidth]{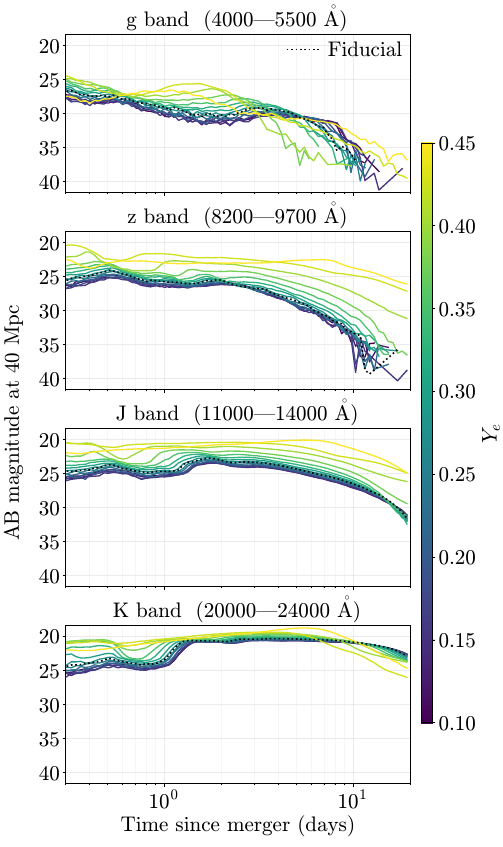}
    \caption{AB-magnitude light curves taken at a distance of 40 Mpc for the electron fraction variation shown in g (4000-5500 \AA), z (8200-9700 \AA), J (11,000-14,000 \AA), and K (20,000-24,000 \AA) bands. The curves are colored by electron fraction. The dotted black line indicates the fiducial model with $Y_e \simeq 0.247$. As shown, increasing $Y_e$ generally produces a brighter transient. Separation between the models occurs and is most pronounced in the red and near-infrared bands. The g-band emission fades rapidly after the first few days. The z, J, and K bands remain comparatively bright for a longer time. This reflects the redistribution of the flux towards longer wavelengths as the ejecta expands and cools. The highest-$Y_e$ models produce the brightest $z$- and $J$-band curves; in the $K$ band they fade below the lower-$Y_e$ models after $\sim$10 days.}
    \label{fig:abmag-ye}
\end{figure}

The light curves in Fig. \ref{fig:abmag-ye} show that the differences in electron fraction models are dependent on the band and epoch. The $g$-band emission fades rapidly after the first several days. The $z$, $J$, and $K$ bands maintain a significant emission for longer. The separation among compositions is most apparent in the red and near-infrared bands. However, the brightness ordering is not strictly monotonic in electron fraction at each epoch. This is due to the measured flux depending simultaneously on the bolometric luminosity, diffusion timescale, and wavelength distribution (which is continuously evolving). 

\subsection{Spectral Redistribution}\label{subsec:yespec}

\begin{figure}
    \centering
        \includegraphics[width=1\linewidth]{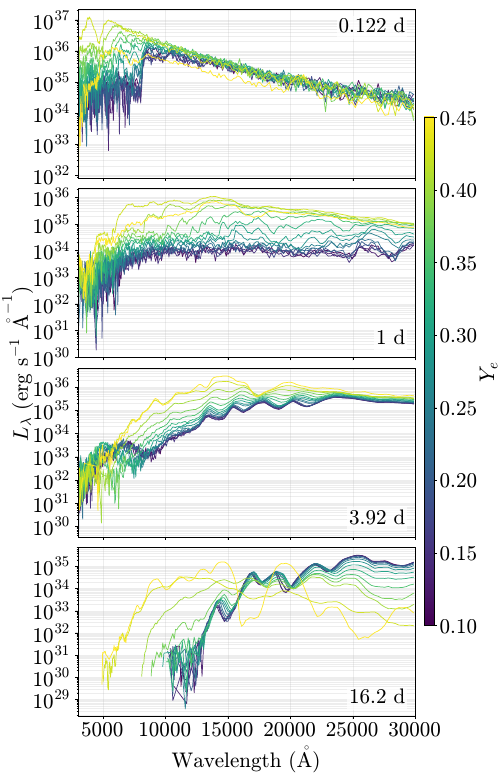}
        \caption{Spectral luminosity for all $Y_e$ models at selected epochs. The curves are colored by initial electron fraction. In early times, luminosity is dominated by short wavelengths for high $Y_e$ models. Over time, kilonova ejecta luminosity shifts redder towards longer wavelengths and the lower $Y_e$ models outshine the more transient higher $Y_e$ models.}
        \label{spectra-dif-epochs}
\end{figure}

Fig. \ref{spectra-dif-epochs} shows the emergent spectral luminosity for all fifteen models at eight representative epochs. During the first day, the high-$Y_e$ runs are generally brighter over most of the optical and near-infrared range. The spectra also show a systematic reddening evolution with time as the shorter wavelength luminosity declines and a larger fraction of the energy appears in the red and near infrared. 

The separation among the runs is largest during the early evolution. At around $0.1 - 0.5$ days, the differences can span orders of magnitude at fixed wavelength. By approximately $1-4$ days, the ordering becomes more complex. Some low electron fraction runs stay strongly suppressed in the optical wavelengths. At later times, the spectra are increasingly dominated by infrared emission. The relationship between local spectral luminosity and initial $Y_e$ is no longer uniform across wavelength. The strong narrow downward trend in the faintest portions of the spectra is likely due to Monte Carlo noise in the bins carrying very little luminosity. 

The shape is most coherent over the intermediate $Y_e$ range. The lowest and highest $Y_e$ models show irregular behavior after $\sim$ several days. Their departures are likely due to the extreme re-weighting of the tracer strength. 

\subsection{Comparison to Previous Composition Studies}\label{subsec:yecompare}

The nucleosynthetic evolution across our tested models is qualitatively consistent with previous reaction network studies. \citet{Lippuner_2015} found that, generally, ejecta become lanthanide poor for $Y_e \gtrsim 0.22 - 0.30$. The exact boundary in that range depends on the entropy and expansion timescale. They also found that the radioactive heating rate changes less abruptly than the lanthanide abundance across this transition. However, some higher $Y_e$ cases produce substantially reduced heating. The rapid decrease in lanthanide and actinide production in our models coupled with the comparatively similar heating histories from $Y_e = 0.10-0.425$ is consistent with this separation between the compositional and heating responses. 

The radiative trends are also consistent with previous opacity calculations in literature. \citet{Tanaka_2020} found that mean opacities of approximately $20-30\ {\rm cm^2\,g^{-1}}$ for $Y_e\leq0.20$, $3-5\ {\rm cm^2\,g^{-1}}$ for $0.25\leq Y_e\leq0.35$, and $1\ {\rm cm^2\,g^{-1}}$ for $Y_e=0.40$ under representative kilonova conditions. Their radiative transfer calculations associated lanthanide-poor, higher-$Y_e$ ejecta with early blue emission and lanthanide-rich, lower-$Y_e$ models with longer-lasting and redder emission. The greater early optical contribution of our higher-$Y_e$ models and the earlier transition to near-infrared dominance in the lower-$Y_e$ models follow the same qualitative behavior \citep{Kasen_2013, Tanaka_2020}. 

Taken together, the calculations demonstrate that changing the ejecta composition at fixed mass and structure produces a coupled nucleosynthetic and radiative response. Increasing the electron fraction suppresses lanthanide and actinide production, modifies the radioactive heating channels, reduces the heavy element opacity, and shifts a larger fraction of the emission toward earlier and shorter wavelength epochs.

\section{Discussion and Conclusion} \label{sec:conclusions}
\subsection{Nuclear Mass Model Variation Discussion}
By varying the nuclear mass model inputs to our nuclear reaction network \texttt{PRISM} and propagating the resulting abundances and heating rates to our Monte Carlo radiative transfer code \texttt{SuperNu}, we have shown through a 30-shell variable $Y_e$ ejecta model that variations in the observables are significantly smaller than those found in single-$Y_e$ studies which are common in the literature. We have successfully attributed the bulk of the differences at later times to variations in the superheavy element production across mass models driving large disparities between the spontaneous fission heating rates of some models. We additionally pinned the differences in spontaneous fission heating to specific isotopes, with \isotope[254]{Cf} dominating the spontaneous fission heating rate at later times for all mass models besides HFB27, which is dominated by a mixture of superheavy isotopes that it uniquely synthesizes in large quantities. \citet{Holmbeck_2023} found that superheavy elements (defined as $Z\geq104$) and actinides brighten the kilonova at fixed ejecta mass, but when the ejecta masses were rescaled to give equal heating at 1~day, the superheavy-rich HFB case evolved like an actinide-free event roughly ten times as massive, and its light curve narrowed and faded faster at early times. Brightness alone therefore cannot separate nuclear input from ejecta mass. In our study, HFB27 does not fade faster as its spontaneous fission heating is sustained past 15~days by $^{259}$Fm, $^{266}$No, and $^{260}$Fm, while the other models converge to $^{254}$Cf-dominated heating with a common late-time slope. This difference is in the shape of the decline rather than its normalization, so a rescaling of ejecta mass cannot remove it. The contrast likely reflects our setup. \citet{Holmbeck_2023} used a single, highly neutron-rich trajectory, whereas we mass-weight 30 tracers with $\langle Y_e\rangle = 0.247$, which dilutes the superheavy yield. Under these conditions, the superheavy signature appears as a change in late-time near-infrared decline rather than as additional early blue emission.

There are several potential reasons why our spread in peak bolometric luminosity is different than other similar studies. First, mass-weighting over a 30-tracer $Y_e$ distribution averages out the model-dependent shifts in lanthanide production that appear at full strength in a single trajectory, and the low-$Y_e$ tracers that produce superheavy nuclei carry little mass. Second, every model is lanthanide-rich enough for the opacity to saturate, so the spread is set almost entirely by the heating. Finally, we quote the spread at peak (${\sim}4$--$5$~d), when $\beta$ decay dominates and differs only modestly between models, while the largest differences arise later, when $\alpha$ decay and fission dominate and our lower ejecta mass reduces thermalization.

The presence of \isotope[258]{Fm} out to approximately 4 days in our HFB27 simulation indicates some decay chain feeding it, as the half-life of \isotope[258]{Fm} is only $370\:\mu$s. Its abundance tracks its production rate, and the fission energy attributed to \isotope[258]{Fm} is effectively released on the decay timescale of those
precursors. Similar but opposite trends appear with \isotope[259-260]{Fm}, where their effect on spontaneous fission heating increases as time goes on. The same behavior appears for \isotope[267]{Rf}
with a half-life of 150 minutes in HFB31, which still contributes 25\% of the spontaneous fission heating at 30~d. Since the fission rates are shared across models, these model-dependent late-time contributions trace the
lifetimes of unmeasured, theoretically predicted precursors rather than the fissioning nuclides themselves. Looking at the \isotope[258]{Fm} chain, we find that HFB27 predicts the production of the theoretical isotope \isotope[258]{Cf}, and ascribes it an unusually long $\beta$-decay half-life of $\approx9$ hours relative to other models. The model predicts that \isotope[258]{Cf} will $\beta$-decay to \isotope[258]{Es}, and \isotope[258]{Es} will $\beta$-decay to \isotope[258]{Fm}.  Identifying these feeding chains is left to future work, but it suggests that measurements of the decay properties of neutron-rich transfermium nuclei could constrain the late-time fission
heating.

Despite being in the same model family, the HFB27 and HFB31 models exhibit very different results in the simulated observables. HFB27 exhibits the brightest peak bolometric luminosity and stays bright late driven by spontaneous fission, and HFB31 is dim at peak driven by relatively low $\beta$ heating. Both are Skyrme-HFB mass models from the Brussels-Montreal collaboration, and in our setup they use the same fission barriers and the same spontaneous fission rates. They differ in how their energy density functionals are constructed. HFB27 uses a Skyrme functional of standard form with a symmetry coefficient $J = 30$~MeV \citep{HFB27}. HFB31 adds density- and momentum-dependent $t_4$ and $t_5$ terms, adopts $J = 31$~MeV, and fits its pairing to nuclear matter gaps that include self-energy corrections \citep{HFB31}. Both models fit measured masses to within ${\sim}0.5$-$0.6$~MeV, so these choices are only weakly constrained near stability, but Skyrme-HFB
mass predictions with and without the $t_4$/$t_5$ extension diverge strongly for heavy neutron-rich nuclei beyond $N \approx 170$ \citep{Scamps_2021}. These differences enter the observables through two separate channels. The
first is the transfermium region. The model masses set the neutron separation energies and $\beta$-decay $Q$-values along the $r$-process path, and hence how much material reaches $A \approx 258$-$276$ by freeze-out. HFB27 produces
roughly 30 times more $X_{Z\geq100}$ (Fig. \ref{fig: xlanxactmw}) than HFB31, and its fission heating is spread across $^{258,259}$Fm, $^{268}$No, and $^{270,271}$Rf. In HFB31, $^{254}$Cf instead carries 54-63\% of the spontaneous fission heating after 5 days (Tab.~\ref{tab:isotope_contributions}).
Because the spontaneous fission rates are identical between the two models, this difference reflects production rather than fission rates. The second channel is $\beta$-decay heating near peak. HFB31 has the lowest
$\beta$-decay heating rate at $t \approx 4$--$5$~d
and the dimmest peak bolometric luminosity. This points to differences in the bulk of the $r$-process distribution rather than the superheavies, so the two models diverge in both the main $r$-process
pattern and the transfermium region.

Future experiments that aim to further constrain nuclear physics properties of exotic isotopes such as some of those analyzed in this study will be crucial to further reducing the nuclear parameter space of future kilonova observations. In particular, measurements of neutron-rich transfermium masses and spontaneous-fission properties at laboratories like the Facility for Rare Isotope Beams (FRIB) at Michigan State University \citep{aprahamian_2018}, the $N=126$ Factory at Argonne National Laboratory \citep{valverde_2026}, and RIKEN in Japan \citep{Wu_2020} would directly constrain the isotopes in Tab. \ref{tab:isotope_properties} whose fission rates currently rest on theoretical FRLDM barriers.

\subsection{Composition Variation Discussion}

The $Y_e$ variations produce clear differences in the kilonova emission. Across the electron fraction variation suite, the total ejecta mass, velocity grid, branch masses, and radial density profile are held fixed. The $Y_e$ variation changes the composition and the associated radioactive heating rates. Increasing $Y_e$ generally suppresses the production of lanthanides, third $r$-process peak nuclei, and actinides. Across most of the sampled range, this is associated with brighter early emission and a larger optical fraction. Lower $Y_e$ models are redder and transfer a larger fraction of their luminosity  into the near-infrared at earlier times. The ordering is not strictly monotonic due to radioactive heating, opacity, and diffusion responding in tandem to the changing composition.

The Vera C. Rubin Observatory Legacy Survey of Space and Time (LSST) provides an excellent opportunity  to test these composition trends across a significantly larger kilonova sample than is currently present in our data. In particular, gravitational wave event follow-ups allow us to gain a more complete picture of these physical systems \citep{2022_Andreoni}. Our results presented here suggest that repeated multiband photometry during the first few days of the event would help distinguish between lanthanide-poor vs lanthanide-rich ejecta based on the strength of early optical emission. In addition to the relevant filter coverage and timing, complementary near-infrared follow up observations would help track the redistribution of emission out of the optical. This would allow characterization of the neutron-rich components that are faint in Rubin's bands. This is one suggestion for how constraints could be placed on the composition for individual events. Further work is still necessary in order to derive more quantitatively confident results however. Still, comparisons across larger samples could establish how widely ejecta compositions vary between mergers. 

\subsection{Joint Study Discussion}

To place the two uncertainty axes on a common footing, we compare the mass model spread in peak bolometric luminosity to the range spanned by the composition grid. Across the full composition grid, $L_{\rm peak}$ varies by a factor of $2.2$ (Fig.~\ref{lum-vs-peak-time}), while the seven mass models span a factor of $1.45$ (Tab.~\ref{tab:lcstats}). The choice of nuclear mass model therefore accounts for a considerable amount of the entire composition-driven dynamic range in peak brightness.

The comparison is sharpened by the strongly non-linear response of $L_{\rm peak}$ to $Y_e$. Above $Y_e \simeq 0.325$, where lanthanide and actinide production falls off rapidly (Sec.~\ref{sec:compresults}), the
peak luminosity rises steeply, with
$d\log_{10} L_{\rm peak}/dY_e \simeq 3.0$; below $Y_e \simeq 0.30$ it is nearly independent of composition, with $d\log_{10} L_{\rm peak}/dY_e \simeq 0.2$. Our mass-weighted tracer ensemble has $\langle Y_e \rangle = 0.247$ and therefore sits in the insensitive branch. Converting the mass model spread into an equivalent shift in electron fraction gives $\Delta Y_e \simeq 0.053$ if evaluated on the steep branch. Within the insensitive branch, no change in $Y_e$ moves the peak bolometric luminosity by the 45\% that the choice of mass model does; matching it requires moving the ensemble onto the steep branch above $Y_e \simeq 0.325$. In this insensitive regime the nuclear physics uncertainty is not a sub-dominant
correction to the astrophysical one, but the leading source of variation in peak brightness as compared to the composition.

Three main consequences follow from this work for the interpretation of observed kilonovae. First, peak bolometric luminosity is seen as a poor diagnostic of ejecta composition in the lanthanide-rich regime when accounting for nuclear uncertainty. Below $Y_e \simeq 0.30$, the ejecta is already saturated in lanthanides and actinides, the opacity and total heating rate are nearly independent of further neutron enrichment, and $L_{\rm peak}$ responds only weakly to composition. In this regime, the choice of nuclear mass model moves the peak brightness by more than any change in $Y_e$. An ejecta composition inferred from a bolometric light curve using a single fixed mass model therefore carries a nuclear systematic that does not diminish with improved photometry.

Second, the two uncertainty axes imprint on different observables, so their combination breaks the degeneracy that either alone leaves. The electron fraction controls the color evolution, the epoch at which the emission transitions from optical to near-infrared dominance shifts by days across our grid (Fig.~\ref{flux-fractions}). The mass model leaves the blue bands nearly
unchanged, where line blanketing from the mass-weighted lanthanide inventory is common to all models, and separates only redward, where the ordering tracks the bolometric luminosity (Fig.~\ref{nmmmulti}). Color therefore constrains the composition and the red and near-infrared luminosity constrains the nuclear input, and a joint fit to both is required to determine either.

Third, the discriminating measurement is late and red. Models whose spontaneous fission heating is dominated by \isotope[254]{Cf} decline differently at late times than HFB27, where \isotope[259]{Fm}, \isotope[266]{No}, and \isotope[260]{Fm} sustain the heating past 15~days (Tab.~\ref{tab:isotope_contributions}). The signature is the shape of the late-time near-infrared decline rather than its normalization. This prediction is subject to an important caveat, as \texttt{SuperNu} assumes local thermodynamic equilibrium (LTE), which becomes increasingly inaccurate as the ejecta enters the nebular phase after $\sim$1 week  \citep{Hotokezaka_2021, Pognan_2021, Pognan_2022}. Because the late-time luminosity tracks the instantaneous deposited power, the bolometric decline shape should be robust, but the distribution of that power across near-infrared bands is not. Non-LTE radiative transfer with non-LTE opacities \citep[e.g.,][]{Pognan_2023} would allow a more precise band-level prediction. AT2017gfo was detected at $4.5\,\mu$m out to 74 days \citep{Kasliwal_2019}, so late-time infrared coverage of a future event, e.g., with JWST, could test this signature directly.

These conclusions are drawn at fixed ejecta structure. The total ejecta mass and velocity distribution remain the dominant observational unknowns and move the peak luminosity by more than either axis considered here; the comparison above isolates the composition and nuclear inputs from that larger uncertainty rather than superseding it.

Future work will include a non-parametric nuclear mass model as part of the variation. One should also investigate simulations which use a per-zone radiative transfer heating rate. Additionally, scanning a grid of neutron star equations of state in the tracer simulations and combining that with mass model variations would expand the coverage of nuclear physics uncertainty in kilonova modeling. 

\begin{acknowledgments}
This work was supported by the U.S. Department of Energy through the Los Alamos National Laboratory. Los Alamos National Laboratory is operated by Triad National Security, LLC, for the National Nuclear Security Administration of the U.S. Department of Energy (contract No. 89233218CNA000001). The research presented in this article was supported by the Laboratory Directed Research and Development program of Los Alamos National Laboratory under project No. 20240170ER. This research used resources provided by the Los Alamos National Laboratory Institutional Computing Program, which is supported by the U.S. Department of Energy National Nuclear Security Administration under contract No. 89233218CNA000001.
A.R. and A.I. express their gratitude towards the X Computational Physics Division Summer Workshop (CPW) at Los Alamos National Laboratory for providing the opportunity to do this research. A.R. would like to thank Kelsey Lund at the Institute for Nuclear Theory for helpful discussion. A.R. would also like to thank Cody Tanner for helpful discussion and for providing feedback on the manuscript. This document has been assigned LA-UR-26-30541.
\end{acknowledgments}

\software{\texttt{PRISM} \citep{Mumpower_2017, Mumpower_2018, Sprouse_2021}, \texttt{SuperNu} \citep{Wollaeger_2013, Wollaeger_2014}, CoH \citep{Kawano_2016}, BeoH \citep{Lovell_2021}}

\newpage
\FloatBarrier
\appendix

\setlength{\intextsep}{6pt plus 2pt minus 2pt}
\setlength{\textfloatsep}{8pt plus 2pt minus 2pt}
\setlength{\abovecaptionskip}{4pt}
\setlength{\belowcaptionskip}{0pt}

\section{Two-Component Smoothly Broken Power-Law Density Profile Construction}\label{subsec:A_powerlaw}

\begin{figure}[H]
    \centering
    \includegraphics[width=0.5\linewidth]{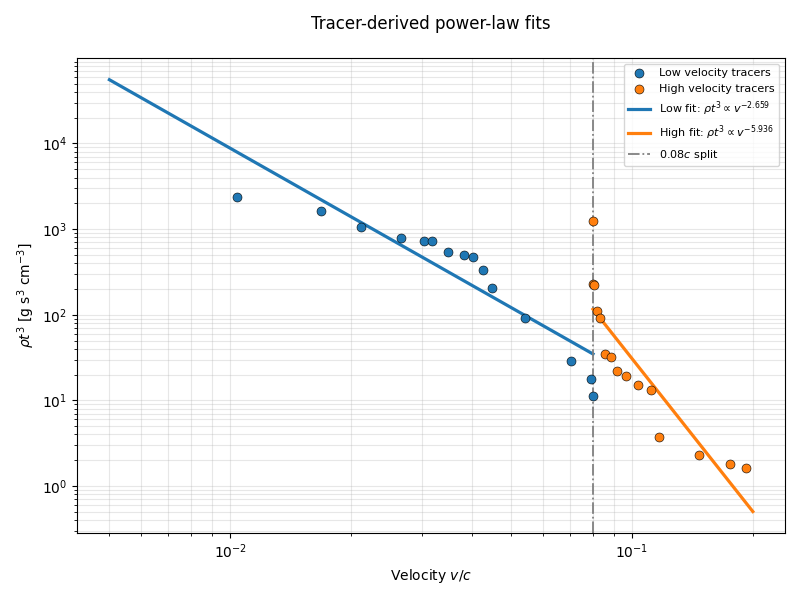}
    \caption{The raw volume-weighted tracer data used to compute the power-law
    for our approximate homologous density profile.}
    \label{sbplfitcompute}
\end{figure}

In this appendix section, we provide more details about the smoothly broken power law density profile used in the study. The tracers of \citet{lund2025neutron} are selected in two radial-velocity
bins, $v < 0.08c$ and $0.08c < v < 0.2c$, containing 66.3\% and 33.7\% of the ejecta mass, respectively. A naive prescription that imposes an independent power law in each bin, each normalized to its own bin mass, conserves mass but generically produces a density discontinuity at the
bin boundary. This discontinuity does not have a hydrodynamic origin as the $0.08c$ boundary is an artificial delineation in the tracer post-processing, not a physical interface in the outflow, and single-component disk winds exhibit smooth density profiles across this velocity range \citep{Tak_2023}. We therefore adopt a smoothly broken power-law (SBPL) in order to create a physically reasonable ejecta density profile, conducive to \texttt{SuperNu} radiative transfer. This SBPL is given by 
\begin{equation}
  \rho(v,t) \;=\; A\,t^{-3}
  \left[
    \left(\frac{v}{v_b}\right)^{s\,n_1} +
    \left(\frac{v}{v_b}\right)^{s\,n_2}
  \right]^{-1/s},
  \label{eq:sbpl}
\end{equation}
with the asymptotic slopes $n_1 = 2.659$ and $n_2 = 5.936$, and the smoothness parameter $s=3$. The tracer-informed slope values $n_1$ and $n_2$ come from computing a log-log regression fit on the ordered $\rho\,t^3$ vs. velocity data from each tracer on each bin (Fig.~\ref{sbplfitcompute}). 
We acquire this fit by noting that the tracer set we use assumes homologous expansion past 127 ms, thus we can use the invariant relationship $\rho\propto t^{-3}$ to compute the relative densities of every tracer in each respective velocity bin that they fall into. Then, motivated by standard kilonova ejecta models, we map the tracer material into a homologously expanding velocity grid in which higher-density material is placed at lower velocity and lower-density material occupies the faster outer layers \citep{Rosswog_2014, Metzger_2019}. In effect, this sorts the tracers by decreasing-density order, restricting the ordering sets to the two velocity bins respectively. The velocity coordinates of this fit come from a relative volume calculation which uses the mass and density that each tracer represents to compute
\begin{equation}
    \Delta V_i\propto\frac{m_i}{(\rho\,t^3)_i},
\end{equation}
where $i$ is a velocity grid cell index.
We then use the homologous volume expansion formula
\begin{equation}
    V(<v)=\frac{4\pi}{3}(v_i^3-v_{i-1}^3)t^3
\end{equation}
to map the tracer volume to velocity space. Thus, the velocity space that a shell occupies is dependent on its density and is given by 
\begin{equation}
    v^3_i-v^3_{i-1}=\frac{3m_i}{4\pi(\rho\,t^3)_i}.
\end{equation} 
This prescription is used to prevent density artifacts from the binning process from contaminating the radiative transfer and thus the observables. The reason we do not use this approach for the final density profile is to avoid numerical issues associated with the high $\rho\,t^3$ shells being razor-thin in volume and to avoid the clear density discontinuity at the bin boundary. Using the power-law fits provided by this calculation (as shown in Fig.~\ref{sbplfitcompute}, computed using the midpoints of the velocity ranges that each tracer occupies) allows us to preserve the tracer-derived density structure through the fitted slopes while ensuring uniform velocity-space sampling, and yields the slopes $n_1=2.659$ and $n_2=5.936$ quoted above.

\begin{figure*}
    \centering
    \includegraphics[width=0.48\textwidth]{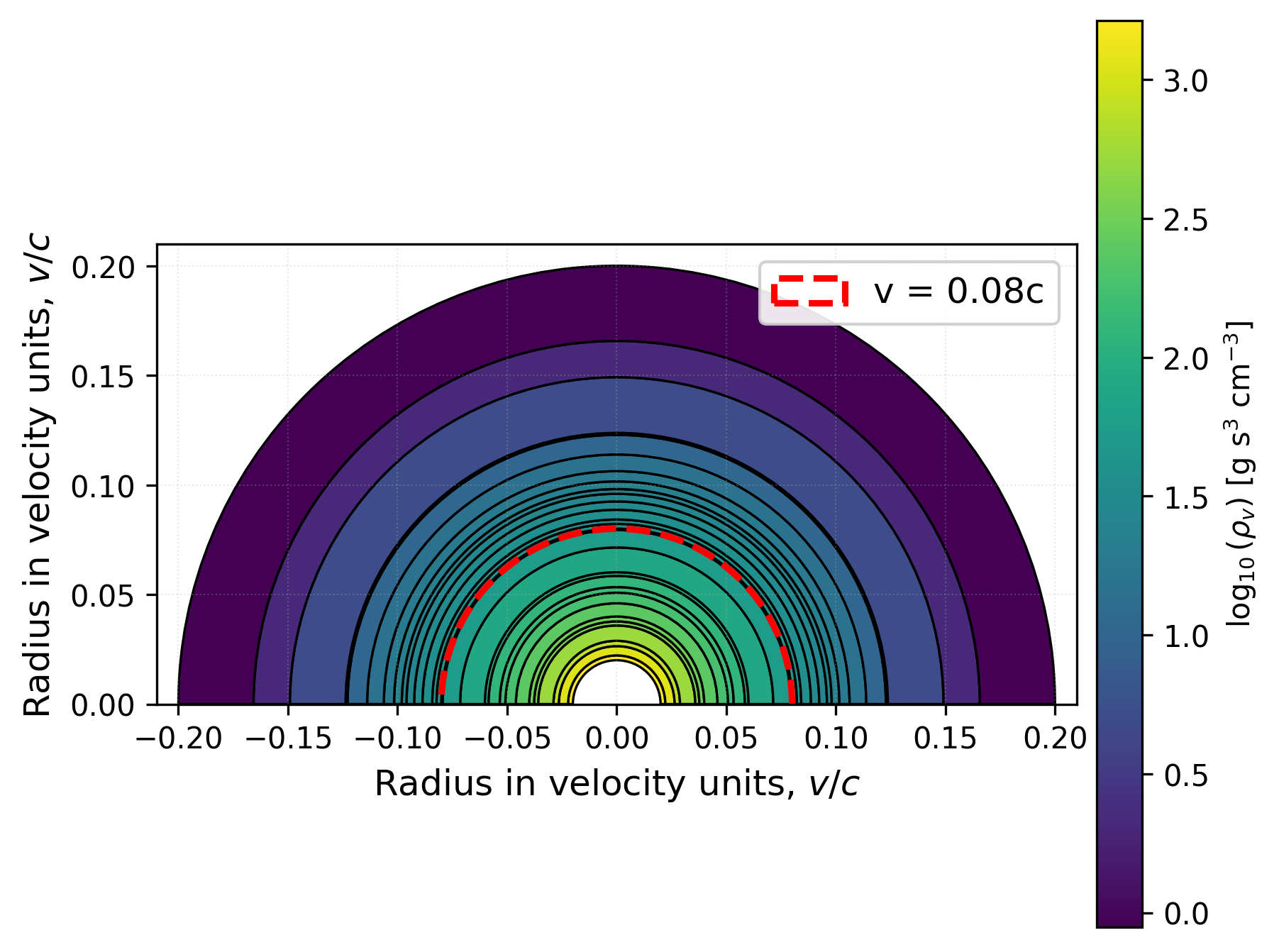}
    \hfill
    \includegraphics[width=0.48\textwidth]{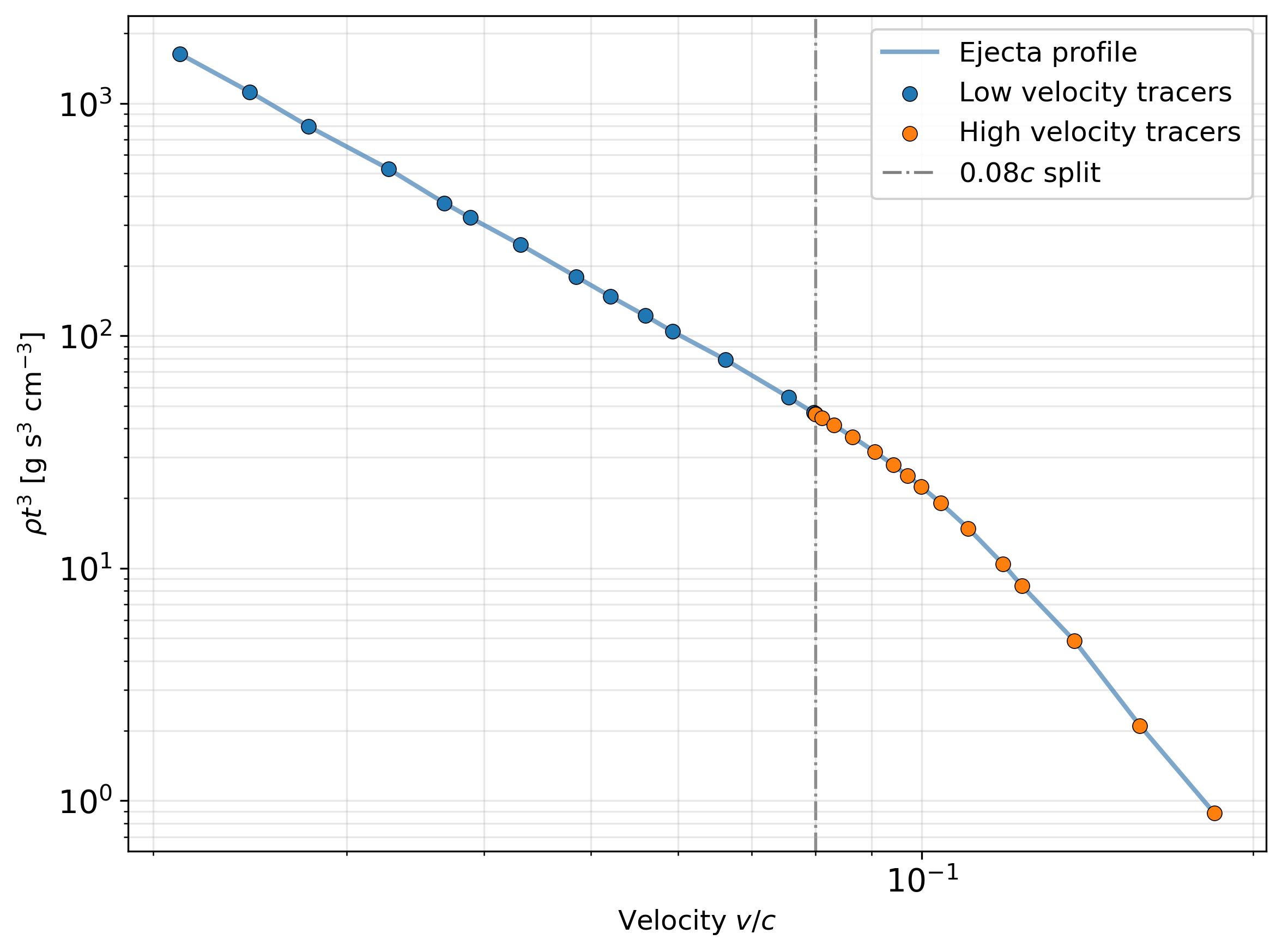}
    \caption{The radial shell density model and the tracer-derived density profile used in the radiative transfer calculations. \textit{Left}: Radial shell model for the density profile used for radiative transfer. \textit{Right}: Density profile for the tracer-derived broken power-law fit data. This profile ensures that each shell has the appropriate density relative to its mass and prevents a density discontinuity at the bin boundary while maintaining a physically reasonable density gradient.}
    \label{fig:density_profiles_combined}
\end{figure*}

The amplitude $A$ is fixed by the total ejecta mass. With a single normalization, per-bin masses can no longer be imposed independently;
that freedom is recovered through the break velocity $v_b$, which we solve by bisection such that the analytic mass fraction interior to $0.08c$ equals the low-bin tracer mass fraction exactly, yielding $v_b = 0.106c$. The selection boundary and the physical density break are thereby decoupled: the profile is continuous in both density and
derivative across $0.08c$, while the imposed mass partition is honored. We now use this SBPL to acquire our density profile that will be used for radiative transfer. Shell boundaries are obtained by inverting the cumulative analytic mass profile, so that each of the 30 shells contains exactly its tracer's mass, with the final low-bin shell edge landing on the $0.08c$ boundary by construction. Within this grid, tracers are assigned radially in order of descending $\rho t^{3}$, evaluated from the homologous tails of the trajectory files ($\rho t^{3}$ is epoch-free, whereas the trajectories' quoted initial densities are defined at differing per-tracer start times and cannot be compared directly). Because angle-averaging erases any recoverable physical one-dimensional ordering, this density-sorted arrangement with the densest material innermost follows the convention of parameterized kilonova modeling under homologous expansion. A model showing the radial shells in velocity space colored by their density is given in the left panel of Fig.~\ref{fig:density_profiles_combined}. The density profile for the final power-law fit is shown in the right panel of Fig.~\ref{fig:density_profiles_combined} (midpoints of the velocity bins plotted). 

Figure~\ref{sbplfitcompute} shows the volume-weighted tracer data underlying
the smoothly broken power-law slopes $n_1$ and $n_2$ quoted above.

\section{Fiducial Velocity-Component Composition}\label{subsec:A_fiducial}

For the composition variations of Subsec.~\ref{subsec:compvar}, the fiducial mass assigned to tracer $(i,c)$ is its tracer mass, $m^0_{i,c}=m_{i,c}$, so that $\sum_{i=1}^{15} m^0_{i,c}=M_c$. The fiducial composition and radioactive heating rate of velocity component $c$ are then the mass-weighted averages
\begin{equation}
\bar{X}^0_{c}(Z,A) = \frac{1}{M_{c}} \sum_{i=1}^{15} m^0_{i,c}\, X_{i,c}(Z,A),
\label{eq:fid_comp}
\end{equation}
\begin{equation}
\bar{\dot{q}}^{\,0}_{c}(t) = \frac{1}{M_{c}} \sum_{i=1}^{15} m^0_{i,c}\, \dot{q}_{i,c}(t),
\label{eq:fid_heat}
\end{equation}
which are Eqs.~(\ref{eq:varied_comp}) and (\ref{eq:varied_heat}) evaluated at $\mu_{i,c}=m^0_{i,c}$.

\section{Nuclear Mass Model Variation Light Curve Data}

\begin{table}[H]
\centering
\caption{FRDM2012-relative bolometric brightness statistics evaluated over the plotted time interval, sorted by decreasing peak luminosity. The columns give the minimum, maximum, and median percent brightness difference relative to
FRDM2012, along with the peak time, peak bolometric luminosity, and percent difference relative to FRDM2012 at peak brightness.}
\label{tab:lcstats}
\small
\setlength{\tabcolsep}{3pt}
\begin{tabular}{lrrrrrr}
\hline
Model & min & max & median & $t_{\rm peak}$ & $L_{\rm peak}$ & peak diff. \\
      & [\%] & [\%] & [\%] & [d] & [erg s$^{-1}$] & [\%] \\
\hline
HFB27    & $+11.36$ & $+119.68$ &  $+37.88$ & 5.08 & $1.552\times10^{40}$ & $+31.64$ \\
DZ33     &  $-0.04$ &  $+39.70$ &  $+16.89$ & 4.76 & $1.335\times10^{40}$ & $+11.51$ \\
FRDM2012 &  $--$    &   $--$    &   $--$    & 4.36 & $1.210\times10^{40}$ &  $--$ \\
SLy4     & $-58.43$ &   $-3.84$ &  $-10.42$ & 4.27 & $1.135\times10^{40}$ &  $-6.14$ \\
UNEDF1   & $-35.65$ &   $+3.05$ &  $-12.44$ & 4.97 & $1.132\times10^{40}$ &  $-4.56$ \\
WS4      & $-15.13$ &   $-3.83$ &   $-8.68$ & 4.56 & $1.121\times10^{40}$ &  $-7.11$ \\
HFB31    & $-16.55$ &   $-6.77$ &  $-11.72$ & 4.46 & $1.074\times10^{40}$ & $-11.08$ \\
\hline
\end{tabular}
\end{table}

Table~\ref{tab:lcstats} collects the FRDM2012-relative bolometric statistics
for each mass model.

\bibliography{refs}{}
\bibliographystyle{aasjournalv7}

\end{document}